\documentclass[aps,prr,floatfix,twocolumn,notitlepage,superscriptaddress,10pt]{revtex4-2}
\usepackage{xcolor}
\usepackage{grffile}
\usepackage{amsmath, amsthm, amssymb, bbold}
\usepackage[normalem]{ulem}
\usepackage{cancel}
\usepackage{bm}
\usepackage{microtype}
\usepackage{hyperref}
\usepackage{setspace}
\usepackage{grffile}
\hypersetup{colorlinks,linkcolor=blue,urlcolor=blue,citecolor=blue}
\usepackage{amsmath}    % need for subequations
\usepackage{amssymb}
\usepackage{graphicx}   % need for figures
\usepackage{verbatim}   % useful for program listings
\usepackage{color}      % use if color is used in text
\usepackage{microtype}
\usepackage[normalem]{ulem}
\usepackage{natbib}
\usepackage{enumitem} 
\usepackage{dsfont} 
\usepackage{hyperref}   % use for hypertext links, including those to external documents and URLs

\newcommand{\bra}[1]{\langle #1|}
\newcommand{\ket}[1]{|#1\rangle}
\newcommand{\average}[1]{\langle #1\rangle}

\newcommand{\dket}[1]{|#1\rangle\!\rangle}
\newcommand{\davg}[1]{\langle\!\langle #1\rangle\!\rangle}

\begin{document}
\title{Nonstabilizerness in open XXZ spin chains: Universal scaling and dynamics}
\author{Doru Sticlet}
\affiliation{National Institute for R\&D of Isotopic and Molecular Technologies, 67-103 Donat, 400293 Cluj-Napoca, Romania}
\author{Bal\'azs D\'ora}
\affiliation{Department of Theoretical Physics, Institute of Physics, Budapest University of Technology and Economics, M\H{u}egyetem rkp.~3, H-1111 Budapest, Hungary}
\affiliation{MTA-BME Lendület "Momentum" Open Quantum Systems Research Group, Institute of Physics, Budapest University of Technology and Economics, Műegyetem rkp. 3., H-1111, Budapest, Hungary}
\author{Dominik Szombathy}
\affiliation{Department of Theoretical Physics, Institute of Physics, Budapest University of Technology and Economics, M\H{u}egyetem rkp.~3, H-1111 Budapest, Hungary}
\affiliation{Nokia Bell Labs, Nokia Solutions and Networks Kft, 1083 Budapest, B\'okay J\'anos u.~36-42, Hungary}
\author{Gergely Zar\'and}
\affiliation{Department of Theoretical Physics, Institute of Physics, Budapest University of Technology and Economics, M\H{u}egyetem rkp.~3, H-1111 Budapest, Hungary}
\affiliation{HUN-REN—BME Quantum Dynamics and Correlations Research Group,
	Budapest University of Technology and Economics, M\H{u}egyetem rkp.~3, H-1111 Budapest, Hungary}
\author{C\u at\u alin Pa\c scu Moca}
\email{mocap@uoradea.ro}
\affiliation{Department of Theoretical Physics, Institute of Physics, Budapest University of Technology and Economics, M\H{u}egyetem rkp.~3, H-1111 Budapest, Hungary}
\affiliation{Department of Physics, University of Oradea, 410087 Oradea, Romania}
\affiliation{MTA-BME Lendület "Momentum" Open Quantum Systems Research Group, Institute of Physics, Budapest University of Technology and Economics, Műegyetem rkp. 3., H-1111, Budapest, Hungary}
	
% \date{\today}
\begin{abstract}Magic, or nonstabilizerness, is a crucial quantum resource, yet its dynamics in open quantum systems remains largely unexplored. 
We investigate magic in the open XXZ spin chain under either boundary gain and loss, or bulk dephasing using the stabilizer R\'enyi entropy $M_2$. 
To enable scalable simulations of large systems, we develop a novel, highly efficient algorithm for computing $M_2$ within the matrix product state formalism while maintaining constant bond dimension—an advancement over existing methods. 
For boundary driving, we uncover universal scaling laws, $M_2(t) \sim t^{1/z}$, linked to the dynamical exponent $z$ for several distinct universality classes.
We also disentangle classical and quantum contributions to magic by introducing a mean-field approximation for magic, thus emphasizing the prominent role of 
quantum-critical fluctuations in nonstabilizerness. 
For bulk dephasing, dissipation can transiently enhance magic before suppressing it and drive it to a nontrivial steady-state value.
These findings position magic as 
a powerful diagnostic tool for probing universality and dynamics in open quantum systems.
\end{abstract}
\maketitle

\section{Introduction} Nonstabilizerness quantifies the deviation of a quantum state from the set of stabilizer states, 
serving as a key resource for quantum advantage~\cite{veitch2014resource, chitambar2019quantum,Leone2024,gidney2024magic} in tasks such as quantum computing and 
simulation~\cite{steane1998quantum,jaeger2007classical,horowitz2019quantum,Oliviero2022}. 
In closed 
systems, unitary evolution generates and propagates magic through coherent interference, with chaotic 
dynamics typically leading to faster magic growth compared to localized 
phases~\cite{vairogs2024extracting, zhang2024quantum, Turkeshi2407.03929, Bejan2024, Niroula2024, paviglianiti2024estimating, ahmadi2024quantifying, bera2025non,Russomanno2025,iannotti2025,jasser2025}. However, in open quantum 
systems, the interplay between coherent dynamics and decoherence fundamentally alters the behavior of magic. 
While dissipation and dephasing generally suppress magic by destroying quantum coherences, 
driven-dissipative systems can exhibit nonequilibrium steady states (NESS) that preserve or even generate 
magic, particularly in the presence of interactions~\cite{aolita2015open,banerjee2018open,Landi2022}. Understanding 
the dynamics of magic in open systems is, therefore, crucial for harnessing quantum resources in realistic, noisy environments~\cite{knill2005, nagata2017,resch2021benchmarking}.  

In this work, we investigate the fate of magic in open quantum systems, focusing on the XXZ spin chain under 
boundary driving and bulk dephasing~\cite{Znidaric2010,Prosen2011,vznidarivc2015, popkov2020exact,mendoza2013heat,heitmann2023}. Using the stabilizer R\'enyi entropy $M_2$ as a measure of magic~\cite{Leone2022}, we 
explore how decoherence and dissipation reshape the Pauli spectrum and influence the growth and saturation 
of magic.

First, we develop a \emph{highly efficient algorithm} for computing the stabilizer R\'enyi entropy 
(magic) in open quantum systems. 
While previous approaches scale as $M^4$ or $M^2$ in the bond dimension $M$~\cite{Lami2023,
tarabunga2024critical,Frau2024}, our method keeps the bond dimension constant, resulting in 
linear memory scaling with $M$ and significantly reduced computational cost. This ensures that 
large bond dimensions, up to $M\sim 1000$, can be handled efficiently in practice. 

With this algorithm in hand, we next investigate the open XXZ spin chain.
For boundary gain and loss, we identify a \emph{universal scaling law for magic} in the XXZ chain, showing $ M_2(t) \sim t^{1/z} $ with  $z$, being the dynamical critical exponent. 
It reflects various transport regimes, including ballistic $(z=1)$, diffusive $(z=2)$, and even the Kardar-Parisi-Zhang (KPZ) dynamics~\cite{ljubotina2019kardar,krajnik2020kardar,scheie2021detection} at the isotropic point  with  $z=3/2$.
Having established the boundary-driven case, we then move to bulk dephasing, which reveals qualitatively different physics and additional insights.
To disentangle classical and quantum contributions to nonstabilizerness, we introduce a mean-field approximation for magic that eliminates quantum correlations while preserving local magnetization contributions. 
This approach provides a classical baseline for magic, allowing us to quantify the role of quantum correlations and identify their impact on transport dynamics and magic generation.

In the presence of bulk dephasing when all sites are coupled to environment through coupling $\gamma_z$, 
we find purely dissipative dynamics with no
coherent evolution at the spin rotational invariant limit. This occurs when starting from a fully polarized state in any direction.
%In this regime, the evolution of the magic density closely follows that of a single qubit under dephasing. 
In general, we find that the time evolution is governed by two distinct timescales: the 
dephasing timescale $\tau_d \sim 1/\gamma_z$, which drives the exponential suppression of magic, and 
the coherence timescale $\tau_c$. The latter is the inverse of the $SU$(2) symmetry-breaking interaction and controls transient oscillations in magic.
Remarkably, dephasing can transiently enhance magic before driving it to zero. 
However, in magnetization-conserving systems, magic does not fully decay but stabilizes at a nontrivial steady-state value. Furthermore, when the dynamics is restricted to the zero magnetization sector, 
starting from certain initial states such as the N\'eel configuration, magic exhibits a robust power-law decay over time, reflecting the underlying gapless nature of the Lindbladian~\cite{Cai2013, SM}.
	
\section{Stabilizer R\'enyi entropy in open systems}
\begin{figure}[t!]
	\begin{center}
		\includegraphics[width=0.7\columnwidth]{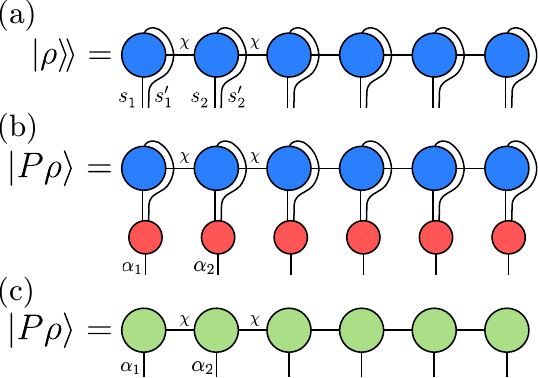}
		\caption{ (a) Illustration of the MPS structure for the vectorized density matrix, with explicit site labels shown.  
			(b) Depiction of the Pauli vector construction and its corresponding MPS structure. 
			Red dots represent the Pauli tensors.  
			(c) After contracting the physical indices, the Pauli vector is expressed as an MPS with the same bond dimension $\chi$ as the original vectorized density matrix.}
		\label{fig:sketch_MPS}
	\end{center}
\end{figure}
In open quantum systems, magic can be extended to mixed states through the stabilizer R\'enyi entropy, defined using the density matrix~\cite{Leone2022}. For a general state $\rho$, the stabilizer R\'enyi entropy $M_2$ is given by  
\begin{equation} \label{eq:m2}  
	M_2 = -\log_2  \frac{ \sum_P |c_P|^4}{\sum_P |c_P|^2},  
\end{equation}
where $P$ represents elements of the $N$-qubit Pauli group~\cite{Poulin2005,Garcia2023,arab2024lecture} and the Pauli coefficients $c_P$ are $c_P = \mathrm{Tr}(\rho P)$. 
In open systems, the Pauli spectrum exhibits both delocalization due to coherent dynamics and decay from nonunitary processes. Even for closed systems, calculating the full Pauli spectrum is computationally expensive for large $N$, due to the exponential growth of the Pauli group. Efficient methods like Monte Carlo sampling~\cite{Tarabunga2023} and tensor networks~\cite{Haug2023,Lami2023,Tarabunga2024} allow for practical evaluation of $M_2$ in closed systems.  

We introduce a new approach to compute magic in open systems, using a vectorized density matrix structure while maintaining a constant bond dimension. 
Details of the algorithm are presented in the Appendix.
% Sec.~\ref{sec:end_matter}}.

This method, extending the framework of Ref.~\cite{Tarabunga2024}, avoids the bond dimension increase required by closed-system algorithms, offering a scalable and computationally efficient tool for quantifying magic in open quantum systems.
	
\begin{figure*}[th!]
	\begin{center}
	\includegraphics[width=\textwidth]{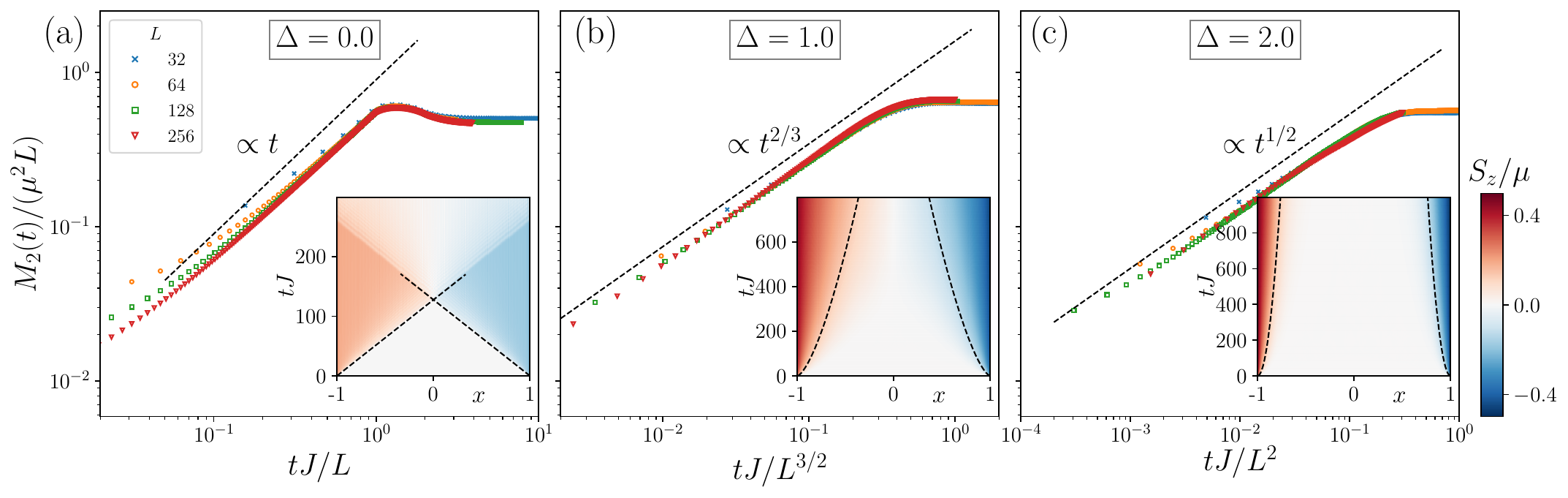}
	\caption{(a,b,c) Time evolution of magic for different anisotropy parameters $\Delta$ and for various system sizes. The insets in each panel present the light-cone formation in magnetization for the respective $\Delta$ as a function of position $x=2j/L-1$, with $j$ numbering the sites.
	The dashed lines in the insets mark the boundaries of the emerging light cones, defined as the points where the magnetization reaches $1/4$ of its maximum value.
	The initial density matrix corresponds to the infinite-temperature state, $\dket{\rho_0} = \dket{\mathbb{1}_\infty}$. 
	For $\Delta=0$, the magic and magnetization light cone exhibit a linear growth, characteristic for the ballistic transport regime.
	At $\Delta=1$, the growth follows a $t^{2/3}$ scaling, indicative of the KPZ universality class, while for $\Delta=2$, it becomes diffusive with a dynamical exponent $z=2$. 
	The legend and color bar are shared for all panels.} 
	\label{fig:Magic_decoherence}
	\end{center}
\end{figure*}

We represent the vectorized density matrix $\dket{\rho}$ as an MPS~\cite{verstraete2004matrix,zwolak2004mixed,daley2004time}, with explicit tracking of the physical index pairs $(s_i, s'_i)$:
\begin{equation}\label{test}
\dket{\rho} = \sum_{s_i, s'_i} A_1^{s_1, s'_1} A_2^{s_2, s'_2} \cdots A_L^{s_L, s'_L} |s_1 s'_1 \dots s_L s'_L\rangle.
\end{equation}
This form allows for efficient manipulation while preserving the system's structure and normalization properties.
To compute the stabilizer R\'enyi entropy, we construct a matrix product operator (MPO) $W$ that encodes the squared action of local Pauli operators according to 
Eq.~\eqref{eq:W}.
The entropy is then given by
\begin{equation}
M_2 = -\log_2 \frac{\average{P\rho |  W^2 |P\rho}}{\average{P\rho | P\rho}}.
\label{eq:M2_MPS}
\end{equation}
Here, $\ket{P\rho}$ is a Pauli-transformed version of $\dket{\rho}$, constructed such that the bond dimension remains unchanged. The structures of $\dket{\rho}$ and $\ket{P\rho}$ are illustrated in Fig.~\ref{fig:sketch_MPS}.

\section{Boundary-driven XXZ spin chain} 
\begin{figure}[t!]
	\includegraphics[width=0.9\columnwidth]{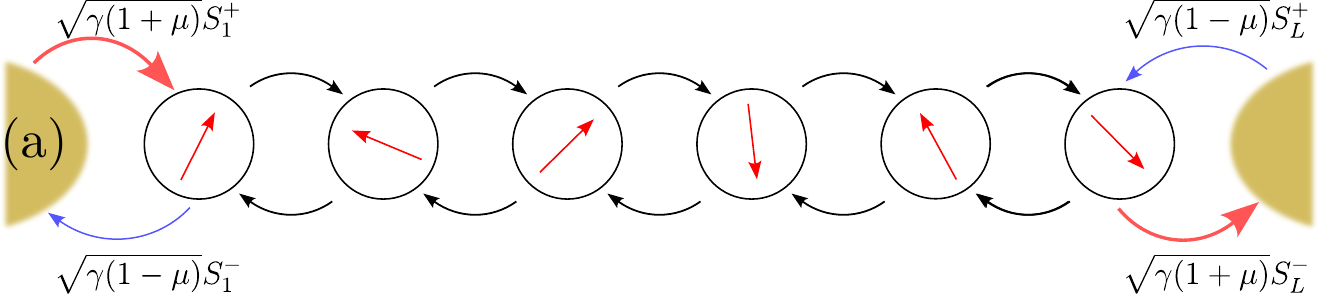}
	\includegraphics[width=\columnwidth]{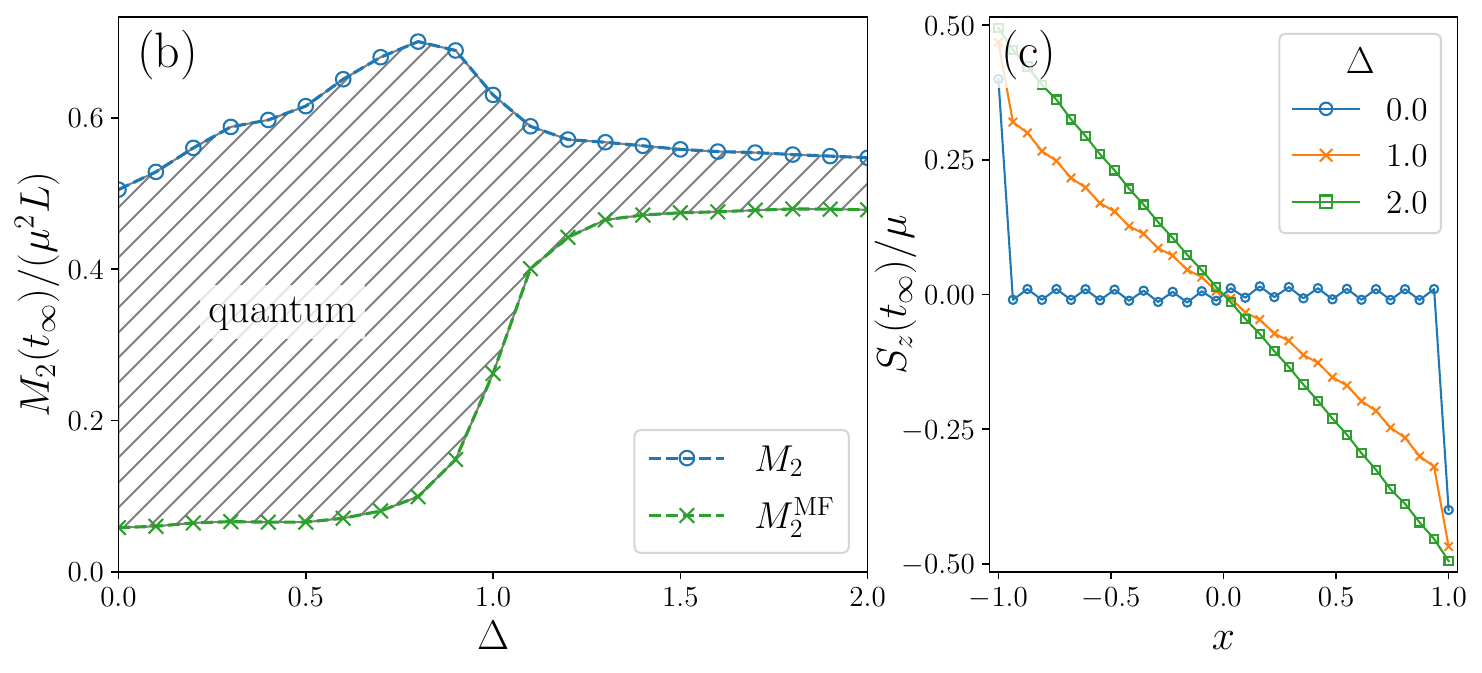}
	\caption{(a) Sketch of the boundary driven XXZ spin chain in the infinite temperature limit. (b) The steady-state values of the total magic, $M_2(t_\infty)$, and the mean-field magic, $M_2^{\rm{MF}}(t_\infty)$, as functions of $\Delta$. The hashed region represents the contribution to magic arising from quantum correlations.
		(c) Steady-state magnetization profile $S_z(t_\infty)$ for three different values of $\Delta$. In both panels, the system size is set to $L=32$.
	}
	\label{fig:magic_saturation}
\end{figure}

	The open XXZ spin chain is a paradigmatic model for studying transport and 
	nonequilibrium steady states in open quantum systems~\cite{Znidaric2011,Prosen2011,Landi2022}. Its Hamiltonian is given by  
	\begin{equation}  
		H = \frac{J}{4}\sum_{j=1}^{L-1}  \left( X_j X_{j+1} + Y_j Y_{j+1} + \Delta Z_j Z_{j+1} \right),  
		\label{eq:H_XXZ}
	\end{equation}  
	where $X_j, Y_j$, and $Z_j$ are Pauli operators at site $j$, and $\Delta$ is the anisotropy parameter.  
	We investigate the generation and depletion of magic in the XXZ chain coupled to external reservoirs. 
	Specifically, we focus on boundary-driven dynamics, where interactions with reservoirs at the chain's edges 
	drive the system into a nonequilibrium steady state~\cite{Znidaric2010,Znidaric2011} as displayed in 
	Fig.~\ref{fig:magic_saturation}(a). 
	The dissipation at the boundaries is described by the Lindblad operators  
	\begin{align}  
		F_{1,1} &= \sqrt{\gamma}\sqrt{1-\mu}\, S_1^-,  & F_{1,2} &= \sqrt{\gamma}\sqrt{1+\mu}\, S_1^+,  \nonumber\\  
		F_{L,1} &= \sqrt{\gamma}\sqrt{1+\mu}\, S_L^-,  & F_{L,2} &= \sqrt{\gamma}\sqrt{1-\mu}\, S_L^+,  
	\end{align}  
	which act at the first and last sites of the chain. The system is initialized in an infinite-temperature 
	density matrix, $\dket{\rho_0} = \dket{\mathbb{1}_\infty}$, a stabilizer state with zero magic. At $t = 0$, 
	the couplings to the reservoirs are switched on, and we track the time evolution of the magic measure $M_2(t)$, 
	along with the local magnetization profile $\langle S_z(x,t) \rangle$. In our numerical simulations, the 
	driving parameter is kept small, $0.01\leq \mu \leq 0.05$.

Following the quench, the system evolves toward a NESS characterized by a steady magnetization current~\cite{Znidaric2011}.
During this evolution, two distinct light-cone structures emerge from the system's boundaries and 
propagate into the bulk. These expanding fronts are clearly visible in the magnetization profiles shown in the insets of Figs.~\ref{fig:Magic_decoherence}~(a)-2(c). 
Their spatial extent $\Delta x$ exhibits a universal scaling $\Delta x \sim t^{1/z}$, reflecting the underlying transport regime: ballistic ($z=1$) in the easy-plane phase, KPZ-type scaling ($z=3/2$) at the 
isotropic point, and diffusive spreading ($z=2$) in the easy-axis regime.
The exponent $z$ dictates the transport behavior, distinguishing different 
universality classes.  A crucial aspect of this evolution is the initial state choice. 
The system is initialized in the infinite-temperature density matrix, $\dket{\rho_0} = \dket{\mathbb{1}_\infty}$, 
which ensures an uncorrelated and homogeneous starting point. This setup allows the observed transport properties 
and the associated dynamical scaling to emerge purely from the interplay between unitary evolution and 
boundary driving, rather than from any pre-existing correlations. In particular, it guarantees that the KPZ 
universality class at $\Delta = 1$ arises intrinsically from the system's dynamics, reinforcing the robustness of the observed scaling behavior.  

For $\Delta < 1$, the system exhibits ballistic transport~\cite{piroli2017transport}, with a linear light cone ($z = 1$). 
At the critical point $\Delta = 1$, the dynamics enters the Kardar-Parisi-Zhang universality class~\cite{scheie2021detection,wei2022quantum,krajnik2023universal}, where transport becomes superdiffusive with a characteristic light-cone scaling $z = 3/2$. 
This anomalous behavior reflects the emergence of strong correlations and fluctuations. 
In contrast, for $\Delta > 1$, transport is diffusive ($z = 2$), resulting in a broader and slower-growing light cone.  
	
Remarkably, the time evolution of magic follows the same universal scaling as the magnetization profile, 
with $M_2(t) \sim t^{1/z}$. Rescaling by $\mu^{2}$ and the system size $L$ reveals a universal collapse of the magic dynamics across different transport regimes, underscoring its role as a robust measure of the system's
universality class. 
In the KPZ regime ($\Delta = 1$), magic exhibits a characteristic $t^{2/3}$ growth, in precise agreement with the expected scaling. 
This highlights magic as a sensitive probe of quantum transport, capturing the emergence of quantum correlations that define the KPZ universality class, beyond what is accessible through conventional observables like magnetization currents.  
Similar to the magnetization profile, magic does not develop uniformly but instead emerges within the two light cones propagating from the system's edges. 
These expanding regions serve as the primary sources of magic generation before it 
eventually saturates in the NESS. As shown in Fig.~\ref{fig:Magic_decoherence} (insets), the transition to the 
steady state coincides with the collision of the two light cones at the center, marking the completion of 
the transport-driven magic buildup. 
	
One of the key findings of this work is the emergence of a universal scaling law for magic, as demonstrated 
in Fig.~\ref{fig:Magic_decoherence}  (a,b,c). Our results establish that the magic follows the scaling form
\begin{equation}  \label{eq:scaling}
	M_2(L, t) \sim L\, f\left(\frac{t}{L^z}\right),  
\end{equation} 
where the scaling function $f(x)$ governs the dynamical evolution across different regimes. 
In the early-time regime ($t \ll L^z$), magic exhibits a universal growth $M_2(t) \sim t^{1/z}$, directly reflecting the system's transport properties, $f(x\ll 1 )\sim x^{1/z}$.
At late times ($t \gg L^z$), magic saturates as $M_2 \sim L$, signaling the approach to the nonequilibrium steady state, which gives $f(x\gg 1)\sim \text{const}$ for the scaling function.
The saturation time $t_0$, at which $M_2$ reaches a plateau, 
	is closely connected to the growth of correlations in the system; it marks the point where the correlation length, constrained by the system size, approaches $\xi \sim L$.
This universal scaling structure provides a unifying framework for understanding its growth across different dynamical universality classes.
\begin{figure}[t!]
	\begin{center}
		\includegraphics[width=0.9\columnwidth]{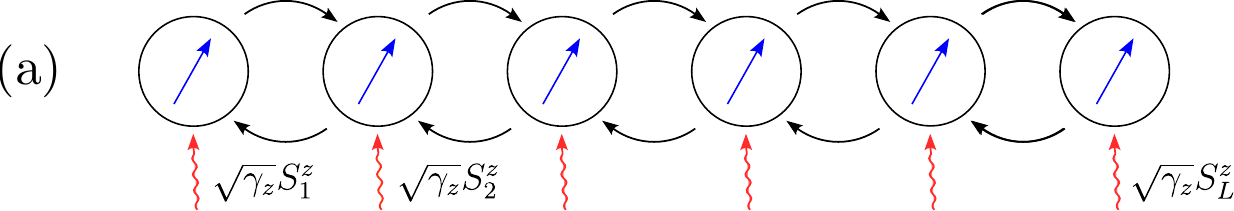}
		\includegraphics[width=\columnwidth]{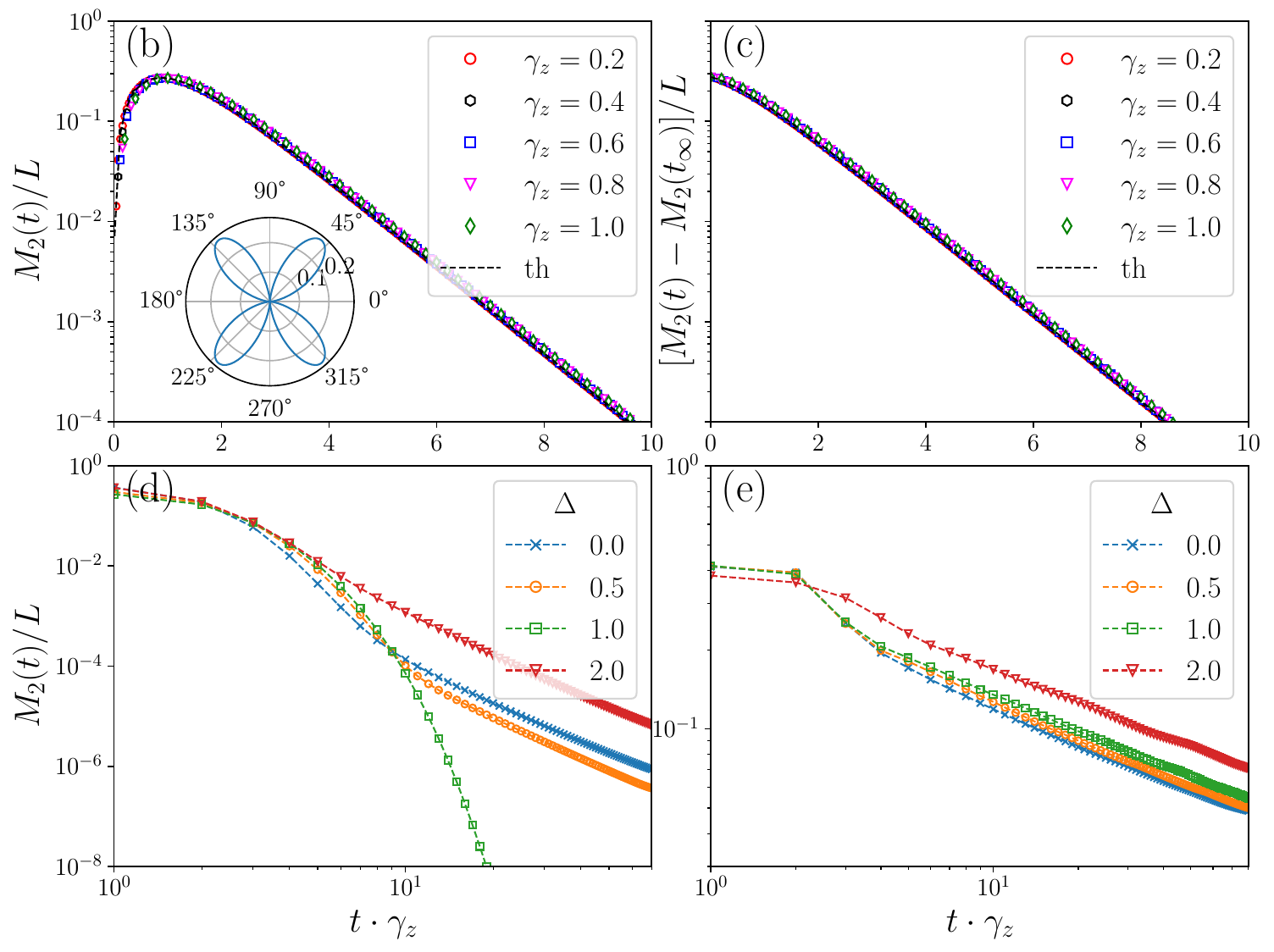}	
		\caption{(a) Sketch of the bulk dephasing modeled with sites coupled to an external reservoir via the jump operator $S^z$.
			(b, c) Time evolution under dephasing affecting all qubits at the isotropic point $\Delta=1$.  
			(b) Starting from an initial density matrix with zero magic, $\rho(0) = (\ket{+}\bra{+})^{\otimes L}$, the magic initially increases before undergoing an exponential decay to zero at late times. The inset displays the magic density in the NESS state $M_2(t_\infty, \theta)$ as a function of the initial polarization angle $\theta$.
			(c) When the initial density matrix is $\rho(0) = (\ket{T}\bra{T})^{\otimes L}$, the system possesses a high degree of magic.
			Due to conservation of the total $S_z$ spin component there is a nonzero magic in the NESS $M_2(t_\infty)=\log_2(6/5)$, subtracted in  (c) from $M_2(t)$. The dashed line denotes the analytical result from Eq.~\eqref{eq:M2_1q}.
			(d, e) Power-law decay of magic $M_2$ for various initial states within the $\langle S_z \rangle = 0$ 
				sector: (d) a product state $\rho(0) = (\ket{+}\bra{+})^{\otimes L}$ and (e) a N\'eel state, 
				both evaluated for $\gamma_z = 1$ and $L = 96$. At the isotropic point $\Delta = 1$, the data in (d) recovers 
				the exponential decay already presented in panel (b).
		}
		\label{fig:Magic_+T}
	\end{center}
\end{figure}
To disentangle classical from quantum contributions, we further introduce a mean-field approximation, which provides a natural baseline for comparison.
\section{Mean-field approximation for magic} 
To quantify the quantum nature of the system, we introduce a mean-field approximation that isolates classical contributions to magic. The 
mean-field density matrix, $\rho_{\rm MF}(t)$, is a product state with bond dimension one, constructed to reproduce the local magnetization 
while discarding all correlations: $\rho_{\rm MF} \approx \bigotimes_{i=1}^{L} \rho_i$, where each local density matrix $\rho_i$ is given by  
$\rho_i = \frac{1}{2} \left( \mathbb{1}_2 + \sum_{\alpha} m_i^{\alpha} \sigma_i^{\alpha} \right)$. Here, $\mathbb{1}_2$ is the identity matrix, $\sigma_i^{\alpha}$ ($\alpha = x, y, z$) are the Pauli matrices at site $i$, and $m_i^{\alpha} = \langle \sigma_i^{\alpha} \rangle$ are the local magnetization components. This formulation allows us to compute the 
mean-field magic $M_2^{\rm{MF}}(t)$ via Eq.~\eqref{eq:m2} and compare it directly with the full magic $M_2(t)$, thereby quantifying the contribution of quantum correlations.  
	
Figure~\ref{fig:magic_saturation}(b) highlights the fundamental role of transport in determining the steady-state magic. In the ballistic 
regime ($\Delta \ll 1$), quantum correlations are dominant, maximizing the gap between $M_2(t_\infty)$ and $M_2^{\rm{MF}}(t_\infty)$.

For small values of $\Delta$ (specifically $\Delta \lesssim 0.6$), the magnetization profile remains largely unchanged [as presented 
in Ref.~\cite{SM}, Fig.~6(b); see also Ref.~\cite{prosen2015matrix}], resulting in an almost constant value of the mean-field magic. 
In this regime, the local magnetization is negligible throughout the bulk and only remains significant near the boundaries. 
Consequently, the mean-field magic $M_2^{\rm{MF}}(t_\infty)$, which depends solely on local expectation values, is primarily 
supported by contributions from the edge sites, leading to its overall smaller magnitude, in contrast to the total $M_2$.
Near $\Delta \approx 1$, both measures of magic rise sharply, signaling enhanced entanglement and a fundamental shift in the 
magnetization profile. 
In the diffusive regime, the gap between full and mean-field magic narrows, reflecting the suppression of quantum correlations. 
In the large-$\Delta$ limit, the steady-state magnetization profile converges to $\langle Z(x)\rangle \sim \arcsin(x)$, with 
vanishing transverse components~\cite{prosen2015matrix}. 
While the  mean-field magic is computed only from local observables, the 
full quantum magic retains contributions from nonlocal correlations such as $\langle\cdots Z_i \cdots Z_j\cdots \rangle$, resulting 
in a finite and persistent gap between the full and mean-field values even as $\Delta \to \infty$, as confirmed by our simulations 
for large $\Delta$.

The explicit decomposition of magic into classical and quantum components is a  
key achievement of this work, as it
provides a direct measure of nonstabilizerness due to quantum correlations. 
Notably, despite being entirely classical, $M_2^{\rm{MF}}(t)$ exhibits the same scaling behavior as the full magic~\cite{SM}, confirming that universal transport 
signatures persist even in the absence of quantum correlations. 
	
\section{Quantum Dephasing}
We investigate bulk dephasing using local Lindblad operators $F_j = \sqrt{\gamma_z} S_j^z$,
	which act independently at each site [see 
Fig.~\ref{fig:Magic_+T}(a)], suppressing quantum coherences in the $S^z$ basis and leading to information loss. To understand its 
effects, we first analyze the single-qubit case. The density matrix evolves as  
$\rho_1(\theta, \varphi, t) = \frac{1}{2}\left[1 + Z\cos\theta + \sin\theta (X\cos\varphi + Y\sin\varphi) e^{-\gamma_z t/2}\right]$, 
where dephasing exponentially suppresses $X$ and $Y$ while preserving $Z$. The corresponding magic follows  
\begin{equation}
	M_2(t) = \log_2 \frac{1 + \cos^2\theta + \sin^2\theta e^{-\gamma_z t}}{1 + \cos^4\theta + \sin^4\theta (\sin^4\varphi + \cos^4\varphi) e^{-2\gamma_z t}}.
	\label{eq:M2_1q}
\end{equation}  
This highlights how dephasing erases quantum coherences while retaining classical information along $Z$. A similar analysis extends 
to two interacting qubits, leading to a general expression for magic (see Ref.~\cite{SM} for details).
	
\section{Heisenberg chain}  
In the Heisenberg limit ($\Delta = 1$), full $SU(2)$ spin symmetry ensures that any fully polarized product state remains an eigenstate, leading to purely dephasing dynamics with no unitary evolution. 
As a result, each qubit evolves independently, and the magic density follows the single-qubit result. 
However, away from this special point, two competing timescales emerge: the dephasing time $\tau_d \sim 1/\gamma_z$, governing the exponential suppression of magic, and the coherence time $\tau_c \sim 1/J|\Delta -1|$, associated with coherent oscillations in magic (see Ref.~\cite{SM} for details). 
At $\Delta = 1$, the absence of coherence-driven dynamics eliminates $\tau_c$, leading to a universal decay dictated solely by $\tau_d$.  
Figure~\ref{fig:Magic_+T} shows the time evolution for two initial states. For $\rho_0 = (\ket{+}\bra{+})^{\otimes L}$ [Fig.~\ref{fig:Magic_+T}(a)], magic starts at zero, peaks at $t \approx 1/\gamma_z$, and then decays exponentially. This transient magic enhancement reflects the interplay between decoherence timescales. When starting from a high-magic state $\rho_0 = (\ket{T}\bra{T})^{\otimes L}$ [Fig.~\ref{fig:Magic_+T}(b)], the conservation of $\langle S_z \rangle$ prevents full decay, stabilizing a nontrivial steady-state magic,  
\begin{equation}
	\lim_{t \to \infty} \frac{M_2(t)}{L} = \log_2 \frac{1 + \cos^2\theta}{1 + \cos^4\theta}.
\end{equation}  
When rescaled by system size, magic dynamics collapse onto a universal curve, matching the single-qubit behavior. The inset of Fig.~\ref{fig:Magic_+T}(a) shows a polar plot of the steady-state magic.  Taken together, these complementary perspectives build a coherent picture of how magic behaves in open quantum systems.

\section{Conclusions}  

In this work, we have uncovered the previously unexplored dynamics of magic in open quantum systems. 
By developing a novel and efficient algorithm for computing the stabilizer R\'enyi entropy $M_2$ in open systems, 
we achieve scalable simulations of large systems under non-unitary dynamics. 
	
For boundary driving, we reveal a universal scaling law for magic, $M_2(t) \sim t^{1/z}$, directly tied to the dynamical exponent $z$ of the underlying transport regime and demonstrate that magic serves as a sensitive probe of quantum transport, including the Kardar-Parisi-Zhang universality class.

We introduce a mean-field approach for magic, which 
isolates the classical contribution to magic through local expectation values.
The difference between the full and mean-field magic serves as a direct probe of nonstabilizerness generated by the many-body correlations built into the system.

For bulk dephasing, we demonstrate that dissipation can 
transiently enhance magic before ultimately driving it toward its steady-state value. Magic exhibits an initial growth, peaks around a characteristic dephasing timescale $t \approx 1/\gamma_z$, and then decays exponentially. 
In systems with conserved magnetization, however, magic is not fully erased and 
stabilizes at a non-trivial steady-state value. In the Heisenberg limit, we derive an analytical 
expression capturing this time dependence. 
Away from the $SU$(2) point, the dynamics are governed by two key timescales: the 
dephasing time $\tau_d \sim 1/\gamma_z$, which controls the exponential decay, and the coherence 
time $\tau_c \sim 1/J|\Delta - 1|$, which accounts for transient oscillations away from the $SU(2)$ 
point. 

Interestingly, when the system is initialized in configurations such as the Néel state or $(|+\rangle\langle +|)^{\otimes L}$, 
both belonging to the zero-magnetization sector, the magic exhibits a power-law decay in time. This behavior reflects the gapless 
character of the Lindbladian dynamics in this regime [see Fig.~\ref{fig:Magic_+T}(d,e) and Ref.~\cite{SM} Sec.~IIc]. 
A more detailed investigation, including the magic dependence on the initial state and system parameters, is the subject of another work.

These results establish magic as a powerful and versatile measure for 
characterizing quantum dynamics in open systems and highlight the complex interplay between unitary and  non-unitary dynamics as well as  symmetry in open quantum systems.
	
	\begin{acknowledgments}
		
This work received financial support from CNCS/CCCDI-UEFISCDI, 
under projects number PN-IV-P1-PCE-2023-0159, 
PN-IV-P1PCE-2023-0987, and PN-IV-P8-8.3-PM-RO-FR-2024-0059, and by the ``Nucleu'' Program within the PNCDI 2022-2027, 
carried out 
with the support of MEC, project no.~27N/03.01.2023, component project code PN 23 24 01 04..
This research was also supported by the National Research, Development and Innovation Office - NKFIH within the Quantum 
Technology National Excellence 	Program (Project No. 2017-1.2.1-NKP-2017-00001), K134437, K142179 by the 
BME-Nanotechnology FIKP grant (BME FIKP-NAT), the 
QuantERA `QuSiED' grant No. 10101773.
We acknowledge the Digital Government Development and Project Management Ltd.~for awarding us access 
to the Komondor HPC facility 
based in Hungary.

	\end{acknowledgments}

	\appendix
	
	\section{ Magic calculation within the vectorized basis} 
	One of the main results of this work is the development of a highly efficient algorithm for computing stabilizer R\'enyi entropy magic in quantum systems, which maintains a constant bond dimension throughout the calculation, ensuring scalability and 
	computational efficiency even for large systems. 
	We represent the vectorized density matrix $ \dket{\rho} $ in matrix product state (MPS) form~\cite{verstraete2004matrix,zwolak2004mixed,daley2004time}
	\begin{equation}
		\dket{\rho} = \sum_{s_i, s'_i} A_1^{s_1, s'_1} A_2^{s_2, s'_2} \cdots A_L^{s_L, s'_L} |s_1 s'_1 \dots s_L s'_L\rangle,
	\end{equation}
	where we explicitly track the pairs of physical indices $ (s_i, s'_i) $ at each site. The tensors $ A_j^{s_j, s'_j} $ are of dimension $ \chi \times \chi $ for $ 1 < j < L $, while the boundary tensors $ A_1^{s_1, s'_1} $ and $ A_L^{s_L, s'_L} $ are a $ 1 \times \chi $ row vector and a $ \chi \times 1 $ column vector, respectively. This structure is illustrated in Fig.~\ref{fig:sketch_MPS}. 
	The normalization condition $ \mathrm{Tr}(\rho) = 1 $ translates to $ \davg{\mathbb{1}_\infty | \rho} = 1 $, where $ \dket{\mathbb{1}_\infty}$ represents the MPS form of the infinite-temperature density matrix.
	To compute the stabilizer R\'enyi entropy, we define the Pauli vector $ \ket{P\rho} $ by applying a tensor product of Pauli operators:
	\begin{equation}
		P_{\bm \alpha} = P_{\alpha_1} \otimes P_{\alpha_2} \otimes \dots \otimes P_{\alpha_L},
	\end{equation}
	and the corresponding vectorized form is
	\begin{equation}
		\ket{P\rho} = \sum_{\alpha_i} B_1^{\alpha_1} B_2^{\alpha_2} \cdots B_L^{\alpha_L} |\alpha_1 \dots \alpha_L\rangle.
		\label{eq:Pauli_vector}
	\end{equation}
	The tensors $ B_j^{\alpha_j} $ are defined as
	\begin{equation}
		B_j^{\alpha_j} = \sum_{s_j, s'_j} \frac{\bra{s_j} P_{\alpha_j} \ket{s'_j}}{\sqrt{2}} A_j^{s_j, s'_j}.
	\end{equation}
	Crucially, the bond dimensions of $ \ket{P\rho} $ are identical to those of $ \dket{\rho} $, ensuring that the computational cost remains manageable. The Pauli vector is normalized such that 
	$\average{P\rho | P\rho} = \sum_{\bm\alpha}\mathrm{Tr}(P_{\bm \alpha} \rho)^2$. 
	Next, we introduce a matrix product operator (MPO) for efficient computation:
	\begin{equation}
		W = \sum_{\alpha_i,\alpha'_i} C_1^{\alpha_1,\alpha_1'} C_2^{\alpha_2,\alpha'_2} \cdots C_L^{\alpha_L,\alpha'_L} |\alpha_1 \dots \alpha_L\rangle\langle \alpha'_1\dots \alpha'_L|.
		\label{eq:W}
	\end{equation}
	The MPO tensors are given by $ C_j^{\alpha_j,\alpha'_j} = B_j^{\alpha_j} \delta_{\alpha_j, \alpha'_j} $, and this construction allows for the computation of the stabilizer R\'enyi entropy (magic) as
	\begin{equation}
		M_2 = -\log_2 \frac{\average{P\rho |  W^2 |P\rho}}{\average{P\rho | P\rho}}.
	\end{equation}
	In practice, we apply  $W$ to $ \ket{P\rho} $ and compute its inner product with itself to obtain the numerator of 
	Eq.~\eqref{eq:M2_MPS}. The constant bond dimension throughout this process is a central feature of the 
	method, making it a powerful and scalable tool for analyzing quantum systems, particularly in the context of stabilizer entropy.

	\bibliography{references}

\end{document}

\title{Nonstabilizerness in open XXZ spin chains: Universal scaling and dynamics}
	\author{Doru Sticlet}
	\affiliation{National Institute for R\&D of Isotopic and Molecular Technologies, 67-103 Donat, 400293 Cluj-Napoca, Romania}
	\author{Bal\'azs D\'ora}
	\affiliation{Department of Theoretical Physics, Institute of Physics, Budapest University of Technology and Economics, M\H{u}egyetem rkp.~3, H-1111 Budapest, Hungary}
	\affiliation{MTA-BME Lendület "Momentum" Open Quantum Systems Research Group, Institute of Physics, Budapest University of Technology and Economics, Műegyetem rkp. 3., H-1111, Budapest, Hungary}
	\author{Dominik Szombathy}
	\affiliation{Department of Theoretical Physics, Institute of Physics, Budapest University of Technology and Economics, M\H{u}egyetem rkp.~3, H-1111 Budapest, Hungary}
	\affiliation{Nokia Bell Labs, Nokia Solutions and Networks Kft, 1083 Budapest, B\'okay J\'anos u.~36-42, Hungary}
	\author{Gergely Zar\'and}
	\affiliation{Department of Theoretical Physics, Institute of Physics, Budapest University of Technology and Economics, M\H{u}egyetem rkp.~3, H-1111 Budapest, Hungary}
	\affiliation{HUN-REN—BME Quantum Dynamics and Correlations Research Group,
		Budapest University of Technology and Economics, M\H{u}egyetem rkp.~3, H-1111 Budapest, Hungary}
	\author{C\u at\u alin Pa\c scu Moca}
	\email{mocap@uoradea.ro}
	\affiliation{Department of Theoretical Physics, Institute of Physics, Budapest University of Technology and Economics, M\H{u}egyetem rkp.~3, H-1111 Budapest, Hungary}
	\affiliation{Department of Physics, University of Oradea, 410087 Oradea, Romania}
	\affiliation{MTA-BME Lendület "Momentum" Open Quantum Systems Research Group, Institute of Physics, Budapest University of Technology and Economics, Műegyetem rkp. 3., H-1111, Budapest, Hungary}
	
\title{Supplemental Material for ``Nonstabilizerness in open XXZ spin chains: Universal scaling and dynamics''}

\begin{abstract}
In this Supplemental Material, we provide additional details and supporting results for the main text. We first describe the 
boundary-driven XXZ spin chain model, highlighting the role of the driving parameter $\mu$ and its impact on nonequilibrium dynamics. 
We then discuss the numerical implementation, focusing on the influence of bond dimension on simulation accuracy. A detailed scaling 
analysis of nonstabilizerness (magic) follows, revealing universal collapse across different transport regimes. Additionally, we 
refine the mean-field approximation to isolate classical contributions to magic, clarifying the role of many-body correlations. These insights further elucidate the intricate interplay between dephasing, coherence, and nonstabilizerness in open quantum systems.
\end{abstract}

\maketitle

\section{Boundary-driven XXZ spin chain}  
The open XXZ spin chain is a paradigmatic model for investigating transport properties and nonequilibrium steady states (NESS) in open quantum systems~\cite{Znidaric2011,Prosen2011,Landi2022}.
Its Hamiltonian is given here for completeness,  
\begin{equation}  
H = \frac{J}{4}\sum_{j=1}^{L-1}  \left( X_j X_{j+1} + Y_j Y_{j+1} + \Delta Z_j Z_{j+1} \right),  
\label{eq:H_XXZ}
\end{equation}  
where $X_j, Y_j$, and $Z_j$ are Pauli operators at site $j$, $J$ the coupling constant, and $\Delta$, the anisotropy parameter controlling the interaction strength in the $Z$ direction.
First we focus on the boundary-driven dynamics, where interactions with reservoirs at the edges of the chain induce a steady-state current, leading to a nonequilibrium steady state (NESS)~\cite{Znidaric2010,Znidaric2011}. 
The dissipation at the boundaries is introduced via Lindblad operators,  
\begin{align}  
F_{1,1} &= \sqrt{\gamma}\sqrt{1-\mu}\, S_1^-,  & F_{1,2} &= \sqrt{\gamma}\sqrt{1+\mu}\, S_1^+,  \nonumber\\  
F_{L,1} &= \sqrt{\gamma}\sqrt{1+\mu}\, S_L^-,  & F_{L,2} &= \sqrt{\gamma}\sqrt{1-\mu}\, S_L^+,  
\end{align}  
which act at the first and last sites of the chain. 
The system is initialized in an infinite-temperature density matrix, $\dket{\rho_0} =  \dket{\mathbb{1}_\infty}$, which describes a stabilizer state with zero magic. 
\begin{figure}[t!]
	\includegraphics[width=0.7\columnwidth]{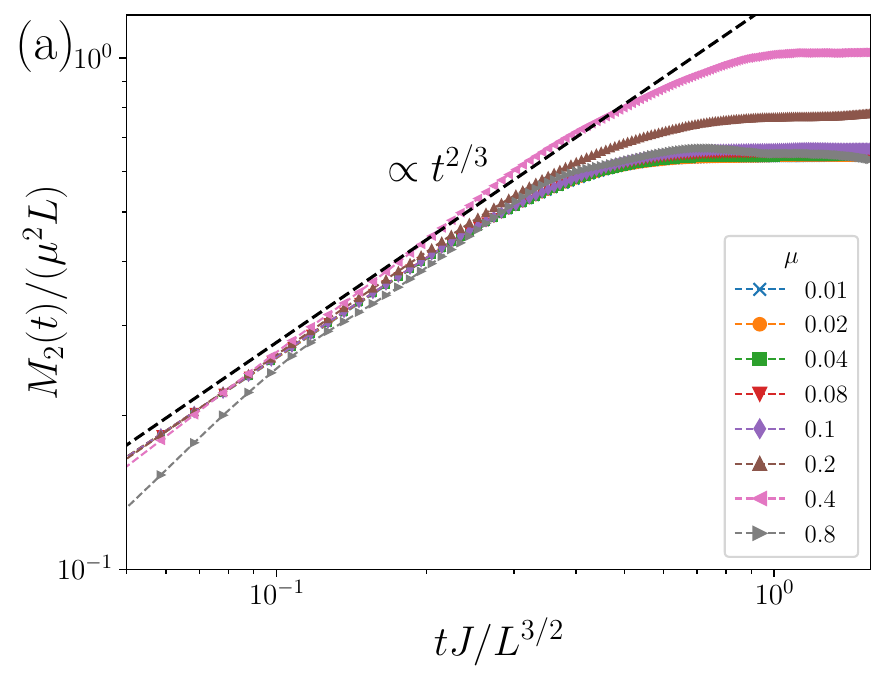}
	\includegraphics[width=0.7\columnwidth]{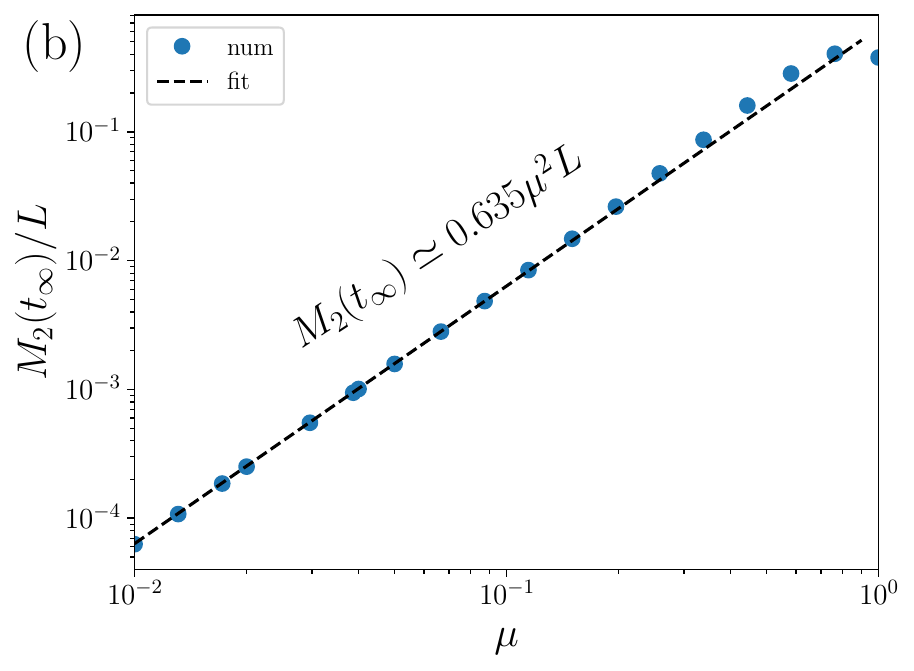}	
	\caption{(a) Time evolution of magic $M_2$ for $\Delta = 1$ at different values of $\mu \lesssim 1$ in a system of size $L = 64$. For $\mu \leq 0.1$, the rescaled magic $M_2 / (\mu^2 L)$ collapses onto a universal curve, indicating a regime 
	of linear response. 
	As $\mu$ increases beyond $\mu^* \approx 0.1$, nonlinear effects become significant, causing deviations from the universal scaling.
	(b) The scaling of magic $M_2$ in the stationary state as a function of the driving parameter $\mu$ shows a quadratic dependence. System size was fixed to $L=32$ and the bond dimension is $\chi=32$.}
	\label{fig:Magic_scaling_mu}
\end{figure}
\subsection{Driving parameter $\mu$}
The driving parameter $\mu$ plays an important role in determining the nonequilibrium dynamics of the boundary-driven XXZ spin chain. It controls the imbalance in the spin injection and extraction at the system's boundaries, thereby setting the strength of the induced magnetization current. 
For small values of $\mu$, the system remains in the linear response regime, where transport properties and the evolution of quantum 
correlations, including magic, exhibit universal scaling behavior. In this regime, the magic  $M_2(t)$ follows a well-defined 
power-law growth, and its rescaled dynamics collapses onto a 
universal curve. 
However, as $\mu$ increases beyond a threshold ($\mu^* \approx 0.1$), 
nonlinear effects start to emerge, altering the transport properties and leading to deviations from universal scaling. 
Therefore, in our numerical simulations, we primarily focus on the weak driving 
regime, $0.01 \leq \mu \leq 0.05$, where the system remains close to linear response. 

As shown in Fig.~\ref{fig:Magic_scaling_mu}(b), the magic $M_2$ in the stationary state demonstrates a distinct quadratic 
dependence on the driving parameter $\mu$ for small values of $\mu$, $M_2(t) \sim \mu^2$. This indicates that the magic remains invariant when rescaled by $\mu^2$, a scaling convention applied consistently throughout our work. This 
quadratic scaling arises as a direct consequence of the linear dependence of the magnetization profile on $\mu$.

\subsection{Role of bond dimension in numerical simulations}  
\begin{figure}[t]
	\includegraphics[width=0.9\columnwidth]{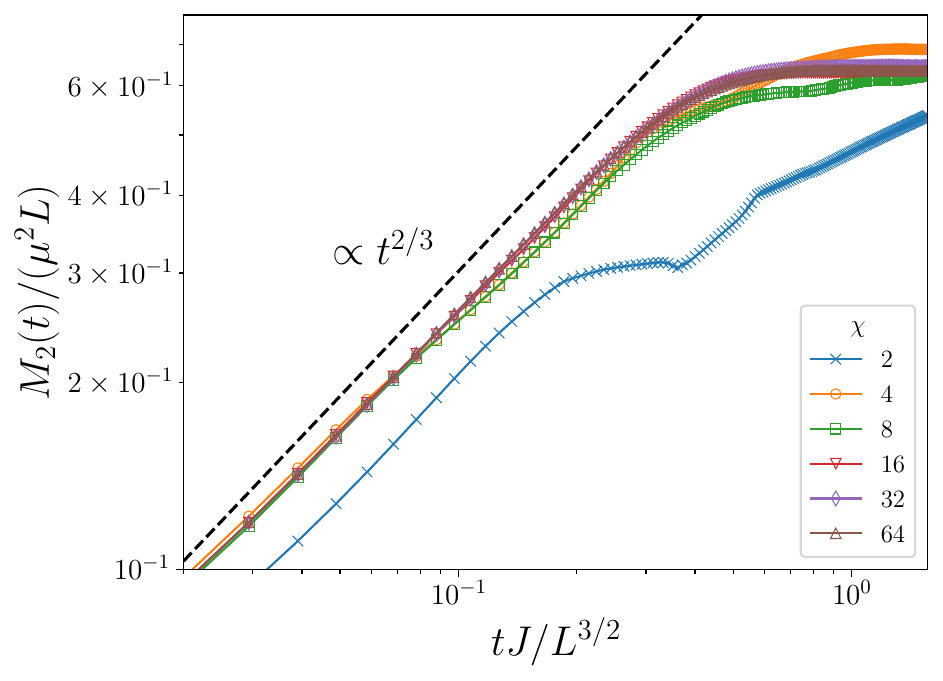}
	\caption{Time evolution of magic for $\Delta = 1$ with different bond dimensions, showing a perfect collapse for bond dimension $\chi \geq 16$. 
	The simulations are performed for a system size of $L = 64$ and a driving parameter of $\mu = 0.05$.}
	\label{fig:bond_scaling}
	\end{figure}

\begin{figure}[tb]
	\begin{center}
	\includegraphics[width=0.86\columnwidth]{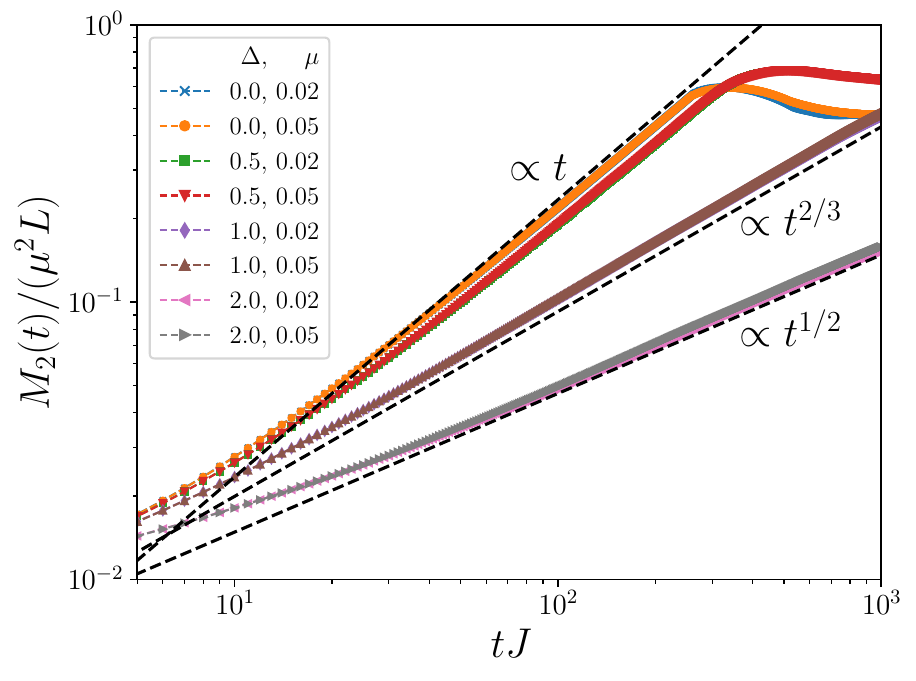}
	\caption{Time evolution of magic in different regimes, exhibiting growth consistent with the dynamical exponents of each universality class: $M_2(t) \sim t^{1/z}$, with $z=1$ for $\Delta<1$ (ballistic), $z=3/2$ for $\Delta=1$ (KPZ), and $z=2$ for $\Delta>1$ (diffusive).} 
	\label{fig:Magic_scaling}
	\end{center}
\end{figure}
\begin{figure*}[tb]
	\includegraphics[width=\textwidth]{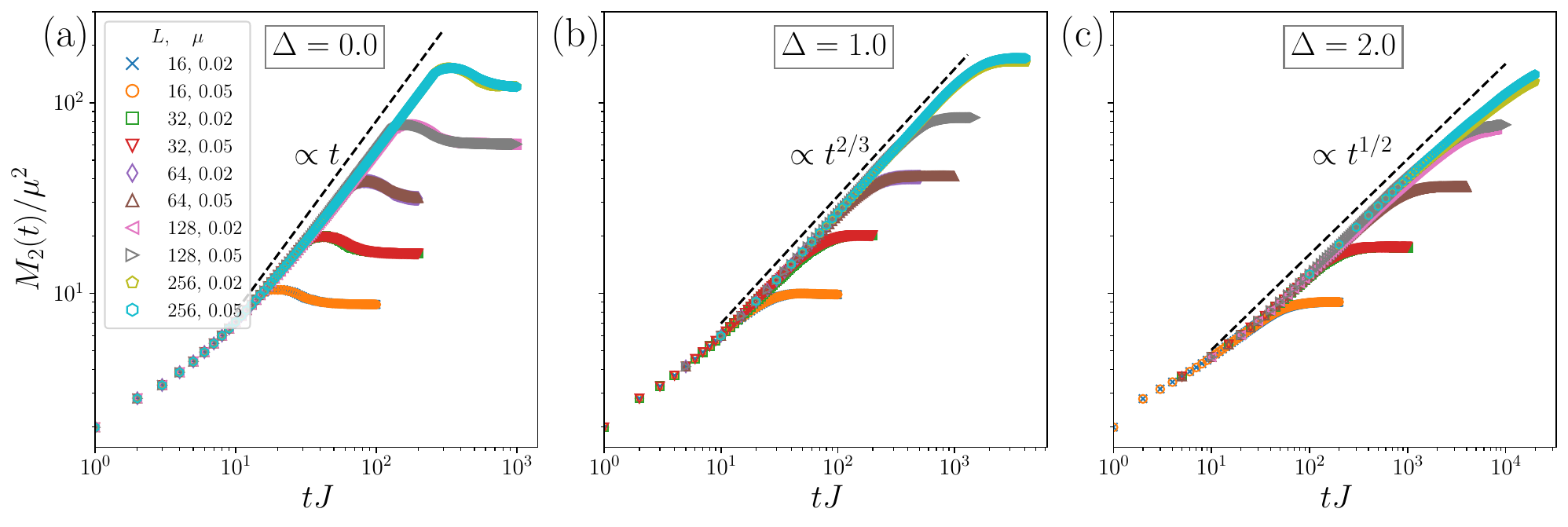}
	\caption{Time evolution of the rescaled magic $M_2 / \mu^2$ for various system sizes and anisotropy parameters $\Delta$. When rescaled with $\mu^2$, the curves for different values of $\mu$ align perfectly, demonstrating a universal collapse. The panels share the legend from (a).
	}
	\label{fig:Magic_scaling_mu_Delta}
\end{figure*}
A key challenge in simulating the dynamics is the efficient representation of the system’s density matrix using tensor network methods. 
In our simulations, we employ the vectorized approach in which the density matrix is represented as a matrix product state (MPS), where the bond dimension $\chi$ controls the accuracy of the representation. 
The required bond dimension depends on both the system size and the nature of 
transport. 
In the ballistic regime ($\Delta < 1$), entanglement growth is relatively moderate, allowing for simulations with a relatively small $\chi$. 
However, in the KPZ regime ($\Delta = 1$), the presence of strong correlations leads to more rapid entanglement growth, necessitating a larger $\chi$ to 
maintain accuracy. 
We systematically check that increasing $\chi$ does not alter the key observables, including the magic measure $M_2(t)$ and the magnetization profiles.  
Figure~\ref{fig:bond_scaling} shows the time evolution of $M_2$ for different maximum bond dimensions, demonstrating that even in the KPZ regime, a relatively small bond dimension ($\chi\geq 16$) is sufficient to accurately capture the dynamics. 
Based on this observation, we usually employ a bond dimension of $\chi = 32$ throughout our simulations.

\subsection{Scaling analysis and universal collapse}  

Following the quench, the system undergoes a dynamical evolution, eventually reaching a steady-state regime characterized by a constant magnetization current~\cite{Znidaric2011}. 
The dynamical exponent $z$ determines the transport behavior, 
distinguishing different universality classes. A crucial aspect of this evolution is the choice of the 
initial state. By initializing the system in the infinite-temperature density matrix, $\dket{\rho_0} = 
\dket{\mathbb{1}_\infty}$, we ensure that the observed transport properties arise purely from the 
interplay between unitary evolution and boundary driving, rather than from any pre-existing 
correlations. 
 When $\Delta < 1$, the system exhibits ballistic 
 transport~\cite{piroli2017transport}, characterized by a linear light cone ($z 
= 1$). At the critical point $\Delta = 1$, transport becomes superdiffusive, characteristic for the KPZ 
universality class~\cite{scheie2021detection,wei2022quantum,krajnik2023universal}, with 
dynamic exponent $z = 3/2$. 
For $\Delta > 1$, transport becomes diffusive ($z = 2$), leading to a broader and more slowly propagating light cone.

An important aspect of our study is the scaling analysis employed to identify universal behavior in the time evolution of magic. 
We observe that magic follows the same universal scaling, given by  
\begin{equation}
M_2(t) \sim t^{1/z},
\end{equation}
as shown in Fig.~\ref{fig:Magic_scaling}. To uncover universal behavior across different transport regimes, we rescale magic 
by $\mu^2$. This rescaling results in a data collapse, demonstrating that the dynamics of magic are governed by the same 
underlying transport mechanisms as conventional observables, such as spin currents. The time evolution of $M_2$ for different 
system sizes $L$ is shown in Fig.~\ref{fig:Magic_scaling_mu_Delta}, which indicates that the stationary value scales with the system size. Therefore, when further rescaled by $L$, a universal 
scaling law, as presented in Eq.~(10) of the main text, emerges for magic. 

\begin{figure}[t!]
	\includegraphics[width=0.7\columnwidth]{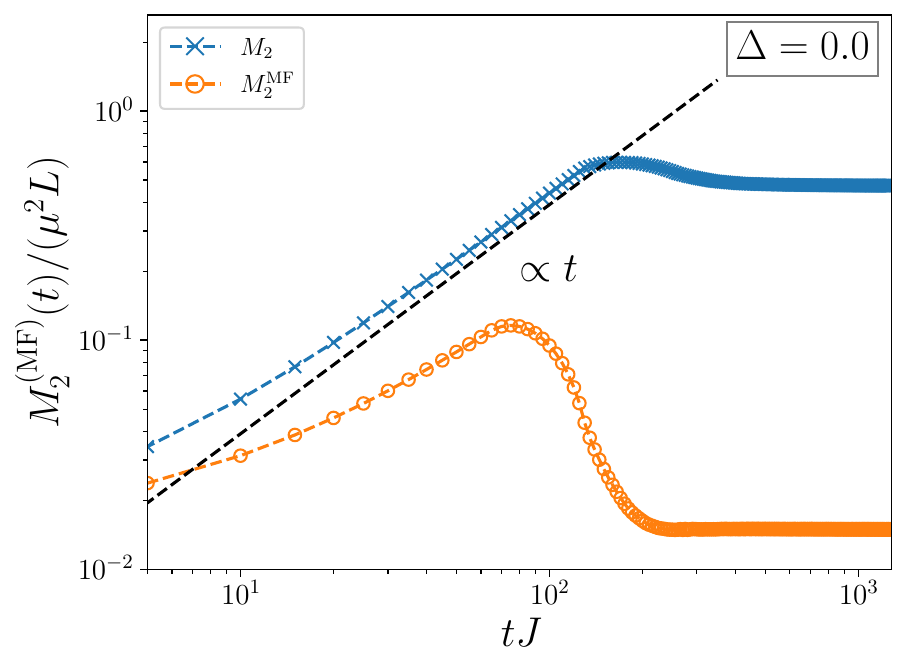}
	\caption{Time evolution of the magic $M_2(t)$ in the XX limit, computed from the full density matrix and the mean-field approximation $M_2^{\rm{MF}}(t)$, which neglects all quantum correlations. [$L=128$ and $\mu=0.05$.]
	}
	\label{fig:magic_comparison}
	\end{figure}
\begin{figure}[t!]
		\includegraphics[width=0.75\columnwidth]{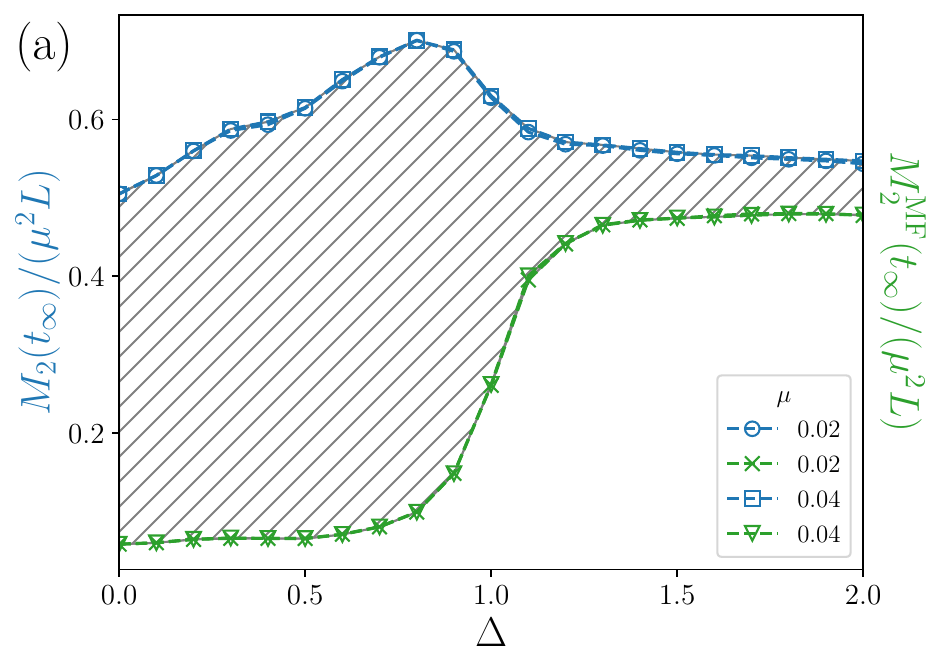}
		\includegraphics[width=0.75\columnwidth]{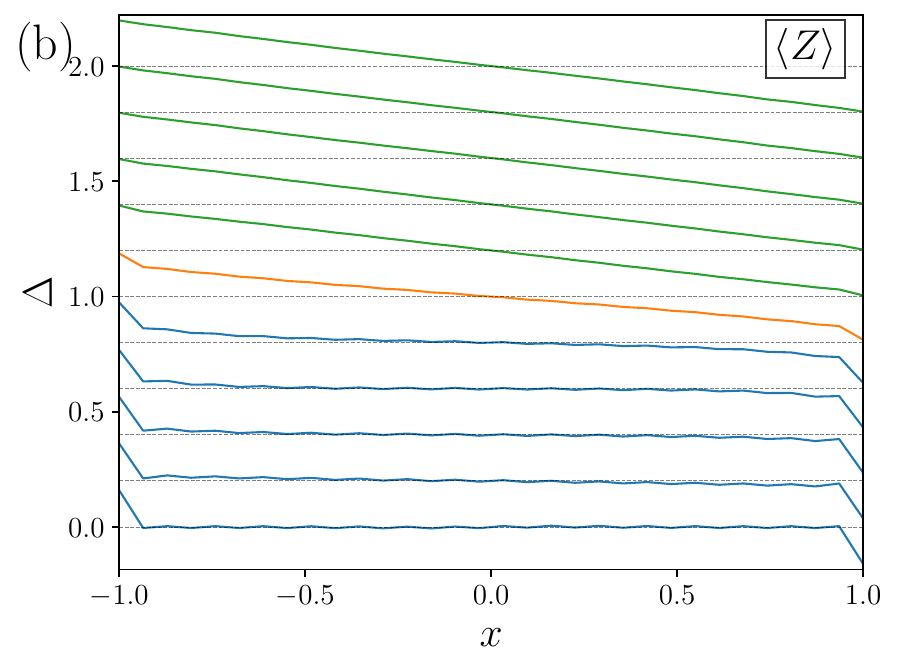}
		\caption{(a) Steady-state values of the full magic $M_2(t)$ (blue) and the mean-field magic $M_2^{\rm{MF}}(t)$ (green) as a function of $\Delta$, for $L=32$ and two values of $\mu$. 
		The hashed region represents the contribution due to quantum correlations.
		(b) Steady-state magnetization profile along the chain for different values of $\Delta$ ranging from 0 to 2 in 
		steps of 0.2 (bottom to top), for $\mu=0.04$ and $L=32$. The magnetization profiles for $\Delta<1$ are in blue, for $\Delta=1$ in orange, and for $\Delta>1$ in green. For $\Delta < 0.6$ and $\Delta > 1.2$, the profiles exhibit small changes.
		The magnetization curves are equally scaled for better visibility and offset vertically for clarity.
		}
		\label{fig:magic_saturation}
\end{figure}

\begin{figure*}[t!]
	\includegraphics[width=\textwidth]{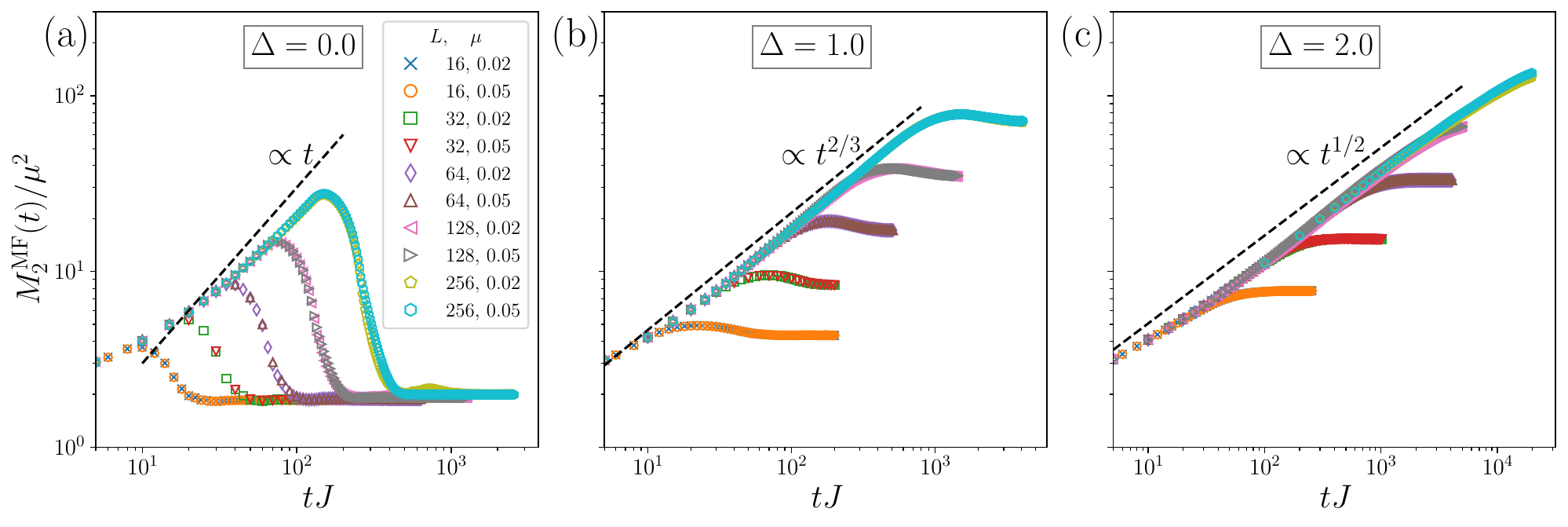}
	\caption{Time evolution of $M_2^{\rm{MF}}(t)$ for various system sizes and driving parameters, highlighting distinct growth behaviors in different transport regimes: linear scaling $M_2^{\rm{MF}}(t) \sim t$ in the ballistic regime ($\Delta < 1$), sublinear growth $M_2^{\rm{MF}}(t) \sim t^{2/3}$ at the $SU(2)$ fixed point ($\Delta = 1$), and diffusive scaling $M_2^{\rm{MF}}(t) \sim t^{1/2}$ for $\Delta > 1$. The panels share the legend from (a).
	}
	\label{fig:magic_classic_scaling}
\end{figure*}
\subsection{Mean-field approximation}
The full density matrix encodes all correlations that emerge following the quench. To assess the quantumness of the state, we employ a mean-field approximation, constructing a mean-field density matrix as a vectorized product state with bond dimension one that reproduces the local magnetization. The procedure consists of the following steps.  
$(i)$ The Lindblad equation is solved in order to obtain the full density matrix $\rho(t)$.  
$(ii)$ The magnetization profile is computed along the chain, $\bm m_{i} = (\avg{X_i}, \avg{Y_i}, \avg{Z_i})$.  
$(iii)$ We define the mean-field density matrix as a product state,
   \begin{eqnarray}
   \rho_{\rm{MF}}(t) &=&\bigotimes_{i=1}^{L}\rho_i(t)\\
   &=&{1\over 2^L}\bigotimes_{i=1}^{L} \left(\mathbb{1}_2 +\avg{X_i}X_i +\avg{Y_i}Y_i +\avg{Z_i}Z_i\right)\notag,
   \label{eq:rho_MF}
   \end{eqnarray}  
which eliminates all correlations while preserving the local magnetization.  
$(iv)$ We use Eq.~\eqref{eq:rho_MF} for the density matrix to compute the mean-field magic  $M_2^{\rm{MF}}(t)$. 
This approximation enables us to compute $M_2^{\rm{MF}}(t)$, where quantum correlations are entirely neglected, and only 
local expectation values contribute when evaluating Pauli strings. 
By comparing $M_2^{\rm{MF}}(t)$ with the full magic $M_2(t)$, we can 
isolate the role of correlations in the system's nonstabilizerness, which implicitly provides us a quantitative measure of the quantumness of the density matrix. 

Figure~\ref{fig:magic_comparison} shows the time evolution of magic $M_2(t)$ in the XX limit ($\Delta = 0$), comparing the full density matrix result, which captures all quantum correlations, with the mean-field result $M_2^{\rm{MF}}(t)$, which neglects them entirely. Since $M_2^{\rm{MF}}(t)$ is fully determined by the magnetization components, its small steady-state value reflects the magnetization profile: in the long-time limit, the central 
region exhibits zero magnetization, while the primary contribution to $M_2^{\rm{MF}}(t)$ originates from boundary sites. 

Figure~\ref{fig:magic_saturation}(a) shows the asymptotic values of magic in the stationary state as a function of the anisotropy parameter $\Delta$. The data highlights how the long-time saturation value $M_2(t_\infty)$, along with its mean-field counterpart $M_2^{\rm{MF}}(t_\infty)$, is influenced by the underlying transport regime. 

For $\Delta \ll 1$, where transport is ballistic, the gap between the full and mean-field magic reaches its maximum, reflecting the strong buildup of quantum correlations. In this regime, the steady-state magnetization profile is almost independent of $\Delta$, resulting in a nearly constant and small mean-field magic, $M_2^{\rm{MF}}(t_\infty) \approx \text{const.}$ for $\Delta \leq 0.6$. This behavior is linked to the fact that for $\Delta < 0.6$, the magnetization profile in the steady state does not depend significantly on $\Delta$, and the primary contribution to magic comes from the boundary regions where the magnetization is finite [see Fig.~\ref{fig:magic_saturation}(b)].

Near the critical point $\Delta \approx 1$, the steady-state value of $M_2(t_\infty)$ is visibly enhanced, reflecting the increased entanglement and quantum complexity in this regime. At the same time, the mean-field magic shows a sharp rise, driven by a significant change in the magnetization profile, with central sites beginning to contribute notably.

In the diffusive regime, the difference between the two magic measures decreases substantially, indicating a suppression of quantum correlations in the steady state.

A crucial finding  is that the scaling analysis of mean-field magic, $M_2^{\rm{MF}}(t)$, reveals the same universal 
behavior as the full magic measure, $M_2(t)$. As shown in Fig.~\ref{fig:magic_classic_scaling}, the time evolution 
of $M_2^{\rm{MF}}(t)$ is precisely governed by the expected dynamical exponents, confirming that it 
captures the universal scaling properties of the system. This result is particularly significant as it 
demonstrates that mean-field magic retains key signatures of quantum transport, despite its classical 
approximation. 
Given that the magnetization profile already exhibits the same scaling behavior, it is natural that 
$M_2^{\rm{MF}}(t)$ follows this scaling.

\section{Quantum dephasing}

We explore the effects of bulk dephasing in the XXZ spin chain, where decoherence is introduced 
through Lindblad operators of the form $F_j = \sqrt{\gamma_z} S_j^z$. These operators suppress 
quantum coherences in the $S^z$ basis, creating a competition between coherent spin-exchange 
dynamics and dissipative relaxation. To gain insight into the many-body behavior, we begin by 
analyzing small system sizes, specifically single- and two-qubit cases, where we derive 
analytical expressions for the time evolution of magic. Building on this, we numerically 
investigate the magic dynamics for larger chains, considering initial stabilizer states with 
varying amounts of nonstabilizerness.
\subsection{Single qubit}

\begin{figure}[t]
	\includegraphics[width=0.9 \columnwidth]{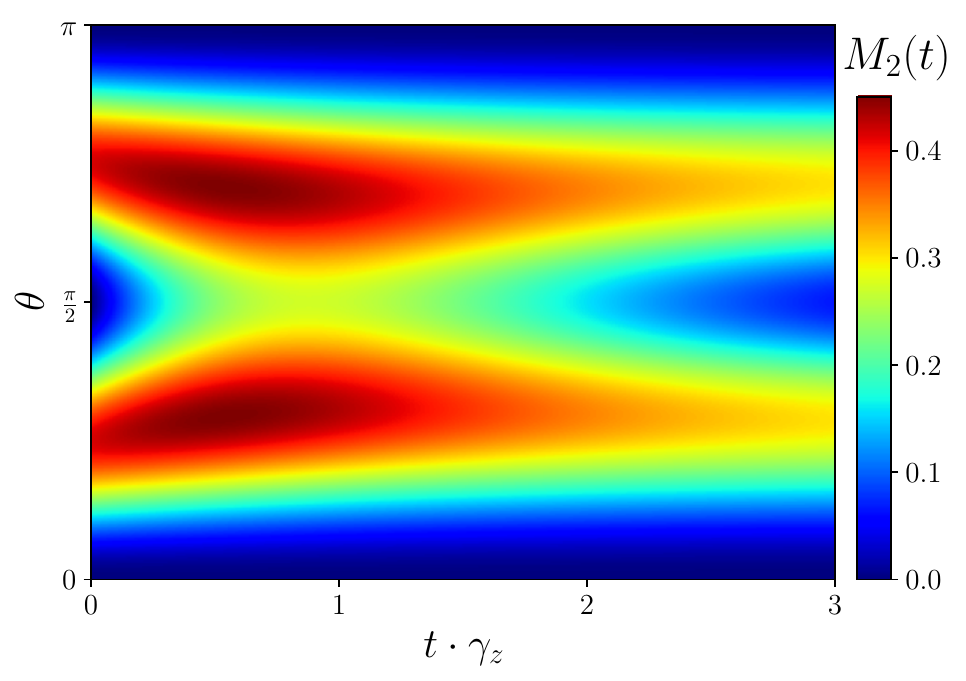}
	\caption{Density plot of $M_2(\theta,t)$ as a function of the initial parametrization angle $\theta$ and time. The initial angle $\varphi$ in Eq.~\eqref{eq:rho_1} is fixed to $\varphi = \pi/2$.}
	\label{fig:q1_map}
	\end{figure}
For a single qubit undergoing dephasing, the Lindblad equation describing the evolution 
of the density matrix can be solved exactly. 
Given an initial state parametrized 
on the Bloch sphere by $\bm n = (\sin\theta\cos\varphi, \sin\theta\sin\varphi, \cos\theta)$,
and defining the Pauli matrices as $\boldsymbol{\sigma} = (X, Y, Z)$, the corresponding density matrix is expressed as  
\begin{gather}  
\rho_1(\theta,\varphi, t=0) = \frac{1}{2} \left( \mathbb{1}_2 + \bm\sigma \cdot \bm n \right).  
\label{eq:rho_1}  
\end{gather}  
Then, the time evolution under pure dephasing takes the form  
\begin{align}  
\rho_1(\theta, \varphi, t) &= \frac{1}{2} \big[ \mathbb{1}_2+ \sin\theta (X\cos\varphi + Y\sin\varphi) e^{-\gamma_z t/2}\notag\\ 
&\quad+ Z\cos\theta\big].  
\end{align}  
By applying the general definition of stabilizer R\'enyi entropy [Eq.~(1) in the main text], the magic is directly computed, yielding Eq.~(11) in the main text, which we reproduce here for completeness,  
\begin{equation}  
M_2( t) = \log_2 \frac{1 + \cos^2\theta + \sin^2\theta e^{-\gamma_z t}}{1 + \cos^4\theta + \sin^4\theta (\sin^4\varphi + \cos^4\varphi) e^{-2\gamma_z t}}.  
\label{eq:M2_1q}  
\end{equation}  
Additionally, if the qubit undergoes unitary evolution under an external magnetic field, the 
Hamiltonian $H = \bm{B} \cdot \bm{S}$ introduces coherent oscillations in the density matrix evolution and implicitly in the magic. In the particular case where $\bm B \parallel \hat{z}$, $S_z$ remains a conserved quantity, and the magic takes a simple 
form that reflects these coherent oscillations,
\begin{widetext}  
\begin{equation}  
M_2(t) = \log_2 \frac{1 + \cos^2\theta + \sin^2\theta e^{-\gamma_z t}}{1 + \cos^4\theta + \sin^4\theta [\sin^4(\varphi + B_z t) + \cos^4(\varphi + B_z t)] e^{-2\gamma_z t}}.  
\end{equation}  
\end{widetext}  
Figure~\ref{fig:q1_map} shows the time evolution of single-qubit magic for different initial density 
matrices~\eqref{eq:rho_1}, parameterized by the angle $\theta$, with a fixed $\varphi = \pi/2$. 
As depicted in the density plot, magic initially increases under pure dephasing, reaches a peak, and 
then decays to a steady-state value. In this regime, the long-time magic $M_2(\theta, t_\infty)$ is 
largely influenced by the initial state's magnetization. While significant magic amplification can 
occur at intermediate times, the steady-state magic always remains lower than its initial value.

\begin{figure*}[tbh]
	\includegraphics[width=0.95\columnwidth]{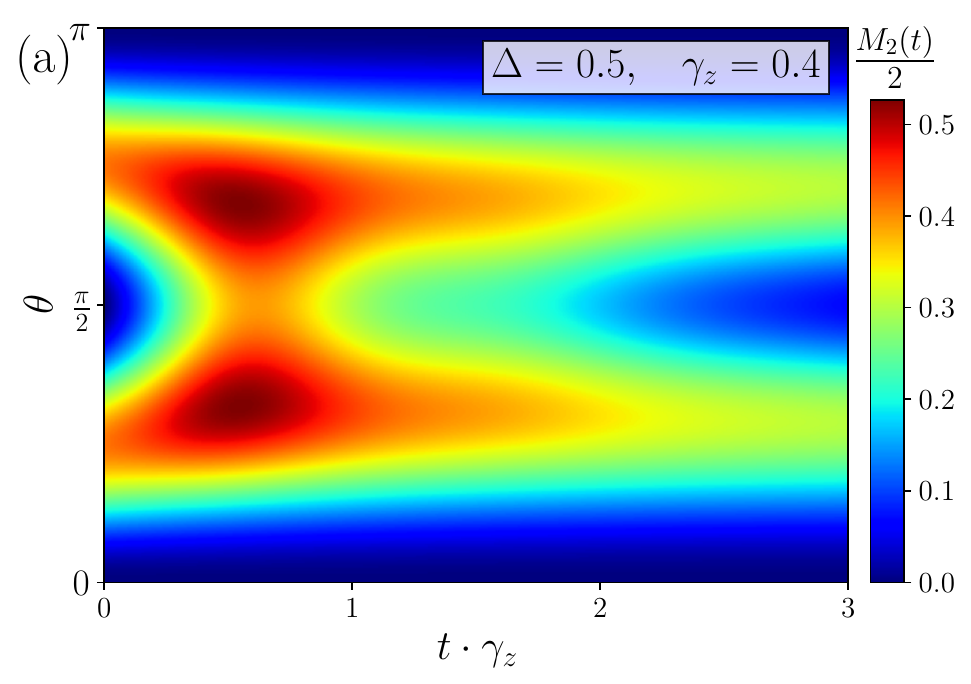}
	\includegraphics[width=0.95\columnwidth]{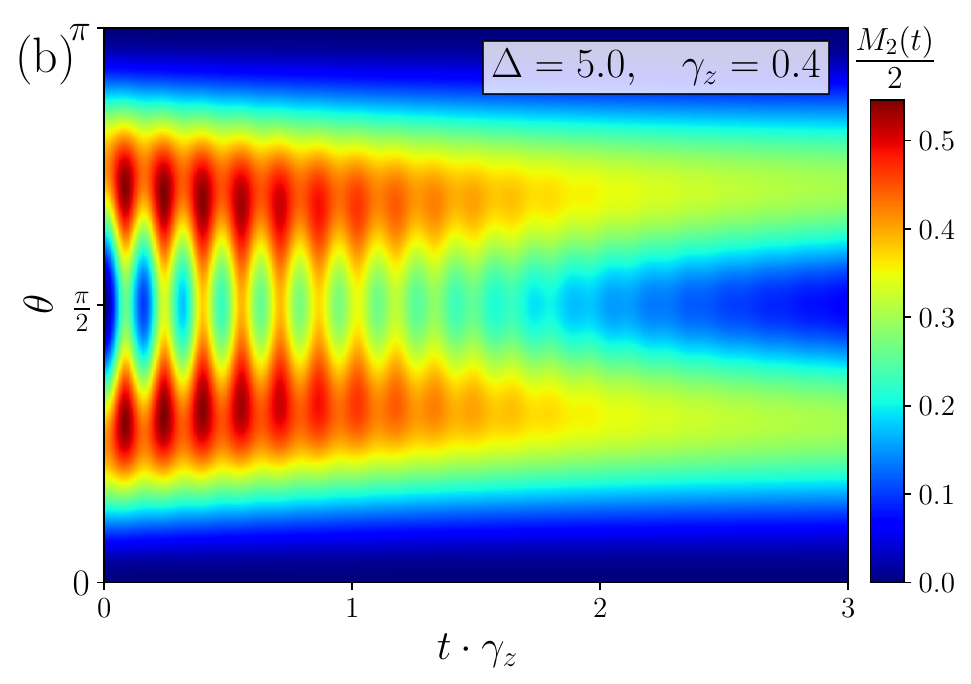}
	\caption{Density plot showing the time evolution for the magic density $M_2(\theta, t)/2$ for two qubits initialized in the same state. 
	Panel (a) corresponds to anisotropy $\Delta = 0.5$, while panel (b) shows the case for $\Delta = 5.0$. In both panels $\gamma_z = 0.4$.}
	\label{fig:q2_map}
	\end{figure*}
\subsection{Two qubits}
\begin{figure*}[ht]
	\includegraphics[width=0.90\textwidth]{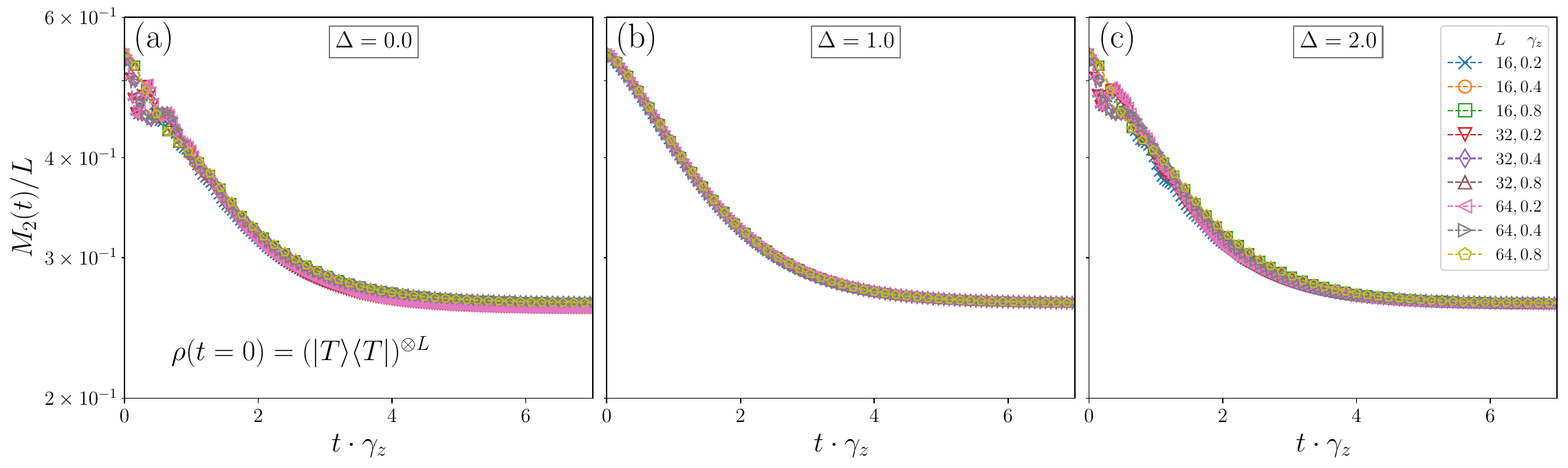}
	\caption{Time evolution of $M_2(t)$ for various anisotropy values, system sizes, and dephasing strengths, as indicated in each panel.}
	\label{fig:magic_dephasing_Delta}
\end{figure*}
\begin{figure*}[tbh]
	\includegraphics[width=0.65\columnwidth]{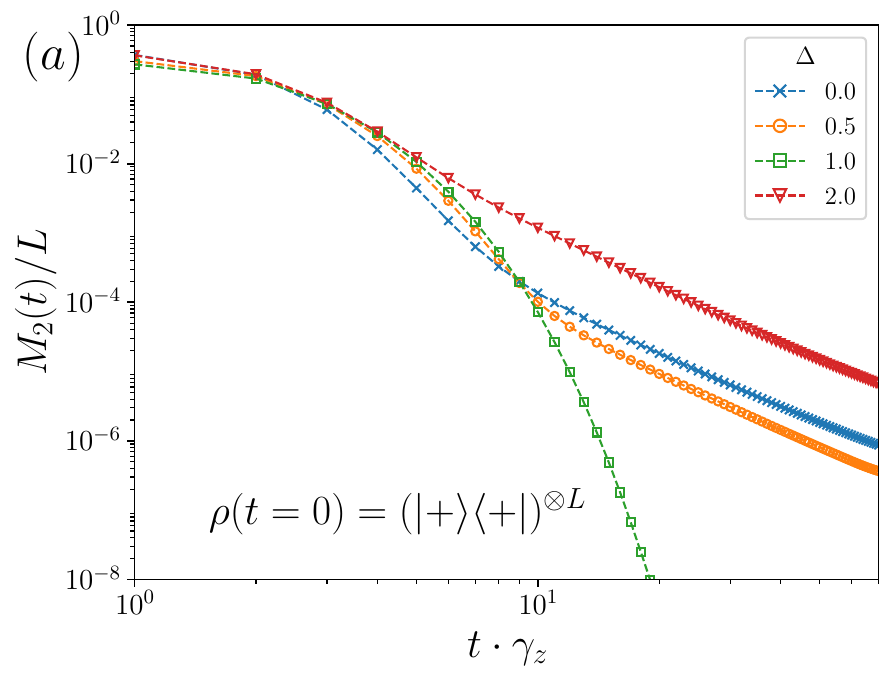}
	\includegraphics[width=0.65\columnwidth]{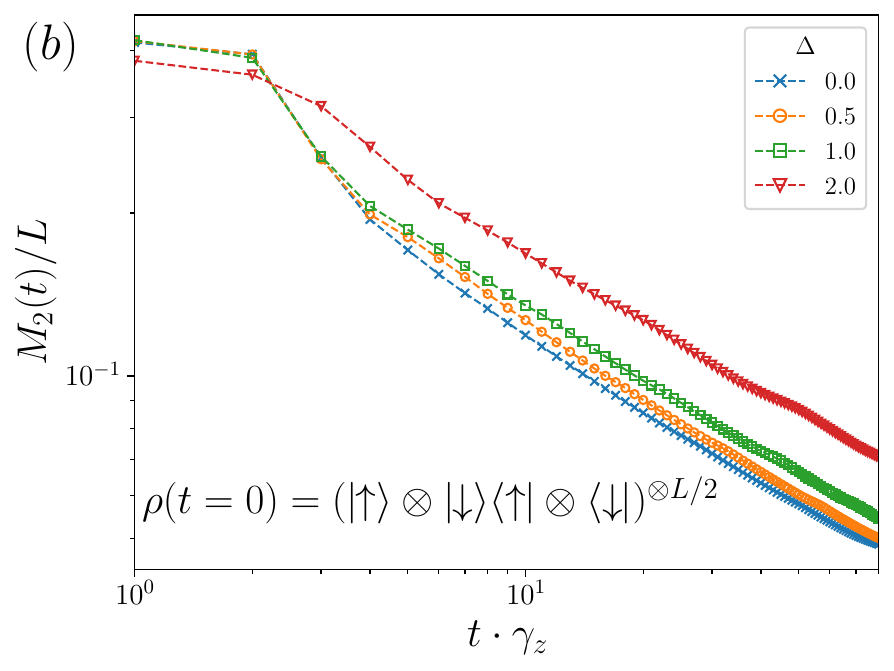}
	\includegraphics[width=0.65\columnwidth]{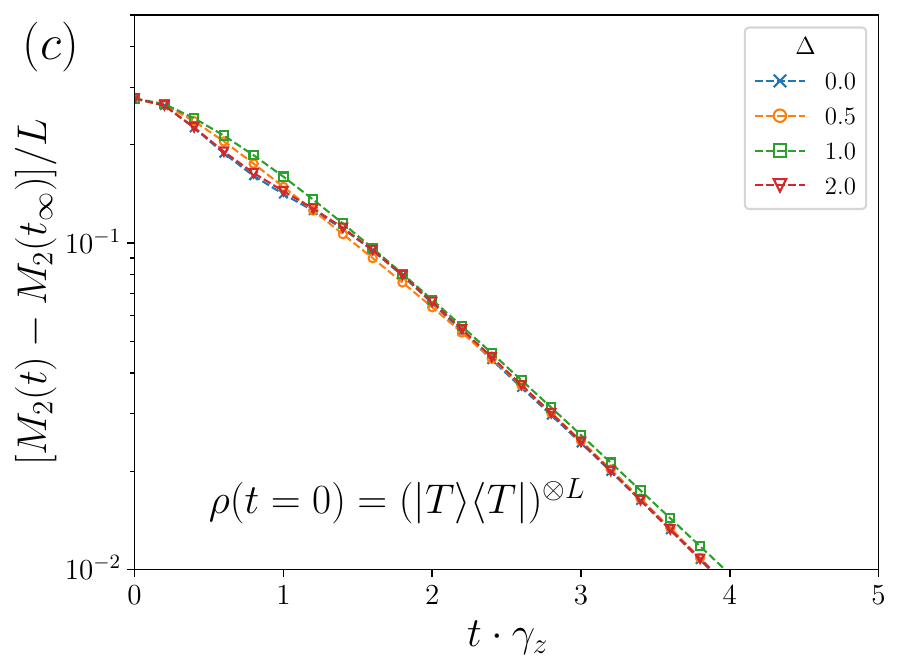}
	\caption{Time evolution of magic $M_2$ for different initial states. (a,b) The evolution starts from the $\langle S_z \rangle = 0$ sector: (a) the product state $(|+\rangle \langle +|)^{\otimes L}$ and (b) the N\'eel state. In (a) and (b), $M_2$ exhibits a power-law decay, $M_2(t) \propto t^{-\alpha}$. 
	There is an exception in (a) at $\Delta = 1$, since the evolution becomes purely dissipative, and the decay transitions to exponential, $M_2(t) \propto e^{-\gamma_z t}$.
	(c) The initial state is $(|T\rangle \langle T|)^{\otimes L}$ and the decay is exponential for all $\Delta$.
	Parameters: system size $L = 96$, dephasing rate $\gamma_z = 1.0$, bond dimension $\chi = 32$.}
	\label{fig:magic_powerlaw}
\end{figure*}

For two coupled qubits interacting via the XXZ Hamiltonian, the time evolution of magic can be determined analytically when the initial state is a product state of identical single-qubit density matrices,  
\begin{equation}  
\rho(\theta, \varphi, t=0) = \rho_1(\theta, \varphi, t=0) \otimes \rho_2(\theta, \varphi, t=0),  
\end{equation}  
where $\rho_j(\theta, \varphi, t=0)$ is given by Eq.~\eqref{eq:rho_1}. 
A particularly insightful case arises in the isotropic limit $\Delta = 1$, where this initial state commutes with the XXZ Hamiltonian due to the presence of $SU(2)$ symmetry. 
Consequently, unitary evolution is completely suppressed, and the system's dynamics is dictated solely by dephasing. 
Since the dephasing operator acts independently on each qubit, the density matrix remains in a product state at all times, and the evolution of magic is fully captured by the single-qubit dynamics.

Beyond this special case, for arbitrary $\Delta$, the Lindblad equation can still be solved exactly for two qubits, allowing for an explicit computation of magic. Summing the contributions from Pauli string expectation values gives the closed-form expression  
\begin{widetext}  
\begin{equation}  \label{eq:M_2q}
M_2(t) = \log_2\frac{(1+\cos^2\theta+\sin^2\theta e^{-\gamma_z t})^2}{(1+\cos^4\theta)^2  
+\sin^8\theta(\cos^4\varphi+\sin^4\varphi)^2e^{-4\gamma_z t}+2\sin^4\theta F(\Delta)e^{-2\gamma_z t}},  
\end{equation}  
\end{widetext}  
where  
\begin{align}  
F(\Delta) &= \re[A]^4 + \im[A]^4 + \re[B]^4 + \im[B]^4, \notag\\  
A &= (\cos\varphi + i\cos\theta\sin\varphi)e^{i(\Delta-1)t}, \notag\\  
B &= (\sin\varphi - i\cos\theta\cos\varphi)e^{i(\Delta-1)t}.  
\end{align}  
At the isotropic point $\Delta = 1$, the magic density, normalized by system size, coincides with that of a single qubit undergoing dephasing. 
This follows directly from taking the $\Delta = 1$ limit in Eq.~\eqref{eq:M_2q}. 
In the long-time limit, magic becomes independent of $\Delta$, and for two qubits, it asymptotically approaches  
\begin{equation}
\lim_{t\to\infty} \frac{M_2(t)}{2} =  
\log_2\frac{1+\cos^2\theta}{1+\cos^4\theta},
\end{equation}  
which reduces to Eq.~\eqref{eq:M2_1q} in the single-qubit case. This result extends to larger systems at 
the $SU(2)$-symmetric point (Heisenberg limit) and is rooted in the fact that fully polarized states (in any direction) 
are eigenstates of the Heisenberg Hamiltonian~\eqref{eq:H_XXZ}. As a result, these states remain unchanged under 
unitary evolution, and the system's dynamics is governed entirely by dissipation. 
Since dephasing acts locally, 
each qubit evolves independently, reproducing the single-qubit result.
Away from the $SU(2)$ point, coherence plays an active role in the dynamics, competing with 
dissipation in shaping the system's evolution.  Figure~\ref{fig:q2_map} presents the time evolution of $M_2(\theta, t)$ for two qubits initially prepared in the same state. In this regime, dephasing leads to an 
exponential decay of magic on a characteristic timescale $\tau_d \sim 1/\gamma_z$, as described in 
Eq.~\eqref{eq:M_2q}. Simultaneously, coherent evolution generates oscillations in magic, characterized by a 
distinct timescale $\tau_c \sim 1/J|\Delta -1|$. These oscillations are particularly pronounced in 
Fig.~\ref{fig:q2_map}(b), becoming dominant when $J|\Delta -1| \gg \gamma_z$. At early times, coherence effects 
drive oscillations in magic, but at longer times, dephasing takes over, leading to an eventual exponential decay 
toward the steady-state value.

\subsection{Numerical results for $L>2$}

As the number of qubits increases, obtaining an analytical expression for magic becomes challenging. However, the key features observed for $L=2$ persist for larger system sizes. We consider the case where all qubits are initially aligned in the same direction,  
\begin{eqnarray}
\rho(\theta,\varphi, t=0) &=& \bigotimes_{i=1}^L\rho_i(\theta, \varphi, t=0)\nonumber\\
&=& \frac{1}{2^L}\bigotimes_{i=1}^L\left(\mathbb{1}_2 + \bm\sigma \cdot \bm n \right),
\end{eqnarray}  
with each spin pointing along the unit vector $\bm{n}$. We analyze how magic evolves under this initial 
condition, and we present numerical results for the time evolution of magic for two different initial states. 

% \begin{figure}[ht]
% 	\includegraphics[width=\columnwidth]{magic_dephasing_T_qc.pdf}
% 	\caption{Characteristic behavior of full magic $M_2$ versus mean-field magic $M_2^{\rm MF}$ for three values of $\Delta$.
% 	At the isotropic point, the two measures coincide, while for $\Delta\neq 1$ $M_2^{\rm MF}$ is less than $M_2$ due to neglecting quantum correlations. The system size is fixed to $L=64$ and $\gamma_z=0.4$.
% 	}
% 	\label{fig:magic_T_qc}
% \end{figure}

We begin by initializing the system in a highly magical state, $\rho(\theta = \pi/4, \varphi = \pi/4, t = 0) = (|T\rangle\langle T|)^{\otimes L}$. 
Figure~\ref{fig:magic_dephasing_Delta} shows the time evolution of $M_2(t)$ for system sizes up to 
$L = 32$. The natural timescale governing the dynamics is the dephasing time, $\tau_d \sim 1/\gamma_z$. 
However, as observed already for $L = 2$, an additional coherence timescale 
$\tau_c \sim 1/J|\Delta - 1|$ introduces oscillations in the early-time dynamics, preventing a 
perfect collapse when rescaling solely by $\tau_d$. Nonetheless, scaling improves at longer times.

Importantly, the conservation of total magnetization $\langle S_z \rangle$ constrains the dynamics to a fixed magnetization sector, meaning the steady-state value of magic is fully determined by the initial $\langle S_z \rangle$. Consequently, the system approaches the same asymptotic value of magic for all anisotropies $\Delta$. Our results show that this steady state is reached exponentially fast, with $M_2(t) - M_2(t_\infty) \propto e^{-\gamma_z t}$.

To investigate the role of this conserved sector more deeply, we analyze dynamics restricted to the $\langle S_z \rangle = 0$ subspace as well.  
Previous work on the XXZ dephasing model~\cite{Cai2013} demonstrated that spin-spin correlations decay as a power law in this 
regime, rather than exponentially, an effect attributed to the gapless nature of the XXZ Lindbladian. 
To test whether magic reflects this behavior, we computed $M_2(t)$ starting from two different 
initial states within the $\langle S_z \rangle = 0$ sector: a Néel state and a uniform product state $(|+\rangle\langle+|)^{\otimes L}$.

Interestingly, when starting from the Néel configuration, magic exhibits a clear power-law decay, 
$M_2(t) \propto t^{-\alpha}$, with an exponent $\alpha \approx 2.66$ [Fig.~\ref
{fig:magic_powerlaw}(b)]. 
For the $(|+\rangle\langle+|)^{\otimes L}$ initial state [Fig.~\ref
{fig:magic_powerlaw}(a)], we observe a slower decay with $\alpha \approx 0.41$, except at the isotropic point $\Delta = 1$, where coherence effects are suppressed and the decay turns exponentially $M_2(t)\propto \exp(-\gamma_z t)$. 
These results suggest that the gapless nature of the Lindbladian manifests 
not only in correlation functions but also in the long-time dynamics of magic.

\bibliography{references}